\documentclass[12pt,preprint]{aastex}

\begin{document}

\shorttitle{Spicules and Jets}
\shortauthors{Mart\'inez-Sykora \& Hansteen \& De Pontieu \& Carlsson}

\title{Spicule-like structures observed in 3D realistic MHD simulations}

\author{Juan Mart\'inez-Sykora$^1$}
\email{j.m.sykora@astro.uio.no}
\author{Viggo Hansteen$^1$} 
\email{viggo.hansteen@astro.uio.no}
\author{Bart De Pontieu$^2$}
\email{bdp@lmsal.com}
\and 
\author{Mats Carlsson$^1$}
\email{m.p.o.carlsson@astro.uio.no}
\affil{$^1$ Institute of Theoretical Astrophysics, University of Oslo, P.O. Box 1029 Blindern, N-0315 Oslo, Norway}
\affil{$^2$ Lockheed Martin Solar \& Astrophysics Lab, Palo Alto, CA94304, USA}

\newcommand{\myemail}{juanms@astro.uio.no}
\newcommand{\viscous}{\underline{\underline{\tau}}}
\newcommand{\resistive}{\underline{\underline{\eta}}}
\newcommand{\komment}[1]{\texttt{#1}}

\begin{abstract}
We analyze features that resemble type {\sc i} spicules 
in two different 3D numerical simulations in which we include horizontal 
magnetic flux emergence in a computational domain spanning the upper 
layers of the convection zone to the lower corona. 
The two simulations
differ mainly in the preexisting ambient magnetic field strength 
and in the properties of the inserted flux tube. We use the {\em Oslo  
Staggered Code} (OSC) to solve the full MHD equations with 
non-grey and non-LTE radiative transfer and thermal conduction along 
the magnetic field lines. We find a multitude of features that show 
a spatiotemporal evolution that is similar to that observed in 
type {\sc i} spicules, which are characterized by 
parabolic height {\it vs.} time profiles, and are dominated by rapid 
upward motion at speeds of 10-30 km~s$^{-1}$, followed by downward 
motion at similar velocities. We measured the parameters of the 
parabolic profile of the spicules and find similar correlations 
between the parameters as those found in  
observations. The values for height (or length) and duration of the 
spicules found in the simulations are more limited in range than 
those in the observations. 
The spicules found in the simulation with higher preexisting ambient 
field have shorter length and smaller velocities. From the simulations, 
it appears that these kinds of spicules can, in principle, be driven by 
a variety of mechanisms that include p-modes, collapsing granules, 
magnetic energy release in the photosphere and lower chromosphere and 
convective buffeting of flux concentrations.
\end{abstract}


\section{Introduction}

The chromosphere is filled with jet-like features, such as dynamic 
fibrils (active regions on the disk), mottles (quiet Sun on the disk) 
and spicules of different types (at the limb). Recently, significant 
improvements in the spatiotemporal resolution of observations with 
Hinode \citep{Tsuneta:2008kc} 
and with the Swedish 1-m Solar Telescope 
(SST) \citep{Scharmer:2008zv} 
to 100~km and a 
few seconds cadence have revealed a complex mix of highly dynamic 
structuring on small spatial scales \citep{de-Pontieu:2007hb,Hansteen+DePontieu2006,De-Pontieu:2007cr}. There has been a long-standing 
discussion on whether and how these different structures are related 
\citep{Beckers:1968qe}.

At the limb, several different types of thin, elongated features, or 
spicules, are observed in H$\alpha$ and other chromospheric 
lines. They are characterized by dynamics on different timescales, 
with so-called type~{\sc i} spicules occurring on timescales of order 
3-10 minutes, and type~{\sc ii} spicules having lifetimes that are 
typically less than 100 seconds. The first type develops speeds of 
10-30~km~s$^{-1}$, reaches heights of 2-9~Mm \citep{Beckers:1968qe} and 
typically involves upward motion followed by downward motion. The 
second type of spicules are more violent with speeds of order 50-100  
km s$^{-1}$, similar heights, and usually only upward motion  
\citep{de-Pontieu:2007hb}. In this paper we focus on a subset of the 
more slowly developing, first type of spicules. 

Recent studies have suggested a close relationship between these 
type~{\sc i} spicules and dynamic fibrils in active regions, and some 
quiet Sun mottles 
\citep{Hansteen+DePontieu2006,de-Pontieu:2007hb,luc2007}. Dynamic 
fibrils dominate the chromospheric dynamics above plage regions in 
active regions, when observed at disk center. They are characterized 
by up- and downward motion that follows a parabolic path as a function 
of time. Typically, the deceleration in this parabolic path is 
different from solar gravitational acceleration. Similar features, so called dark 
mottles, are seen in and around small flux concentrations of the quiet 
Sun magnetic network. The similarity in dynamic behavior of mottles, 
dynamic fibrils and some type~{\sc i} spicules with up- and downward 
parabolic motion, and significant correlations between parabolic 
parameters such as deceleration and maximum velocity strongly suggest 
that these features are driven by the same mechanism. 
\citet{Hansteen+DePontieu2006} and \citet{De-Pontieu:2007cr} found 
jet-like features with similar properties in self-consistently driven 
2D-MHD simulations. The jet-like features in these simulations were 
found to be caused by shock-waves driven by leakage of convective 
motions and oscillations from the photosphere into the chromosphere 
along magnetic field concentrations. This is the mechanism originally 
suggested by \citet{De-Pontieu:2004hq} to explain the 
quasi-periodicities observed in dynamic fibrils that permeate the 
upper transition region of coronal loops \citep{De-Pontieu:2003ct}. \citet{Heggland:2007jt} used 
a wide variety of 1D simulations to show that a mechanism in which 
shock waves drive these jets can naturally explain the correlations 
between parabolic parameters that was found in the observations and 2D 
simulations. 

In this paper we take the logical next step and analyze similar 
jet-like features that we find in 3D-MHD simulations. These 
simulations are quasi-realistic and include non-grey and non-LTE 
radiative transfer with scattering and thermal conduction along the 
magnetic field lines. The simulations include the upper layers of the 
convection zone up to the lower corona. 
The organization of the paper is as follows: \S~\ref{sec:equations} briefly 
describes the numerical methods, initial conditions and boundaries 
employed in the numerical code. In \S~\ref{sec:st}, we describe the 
correlation between the properties of the 150 spicule-like features that we found 
in the simulations. While all of these spicule-like features seem to be driven 
by magnetoacoustic shocks, we find that a variety of events can 
produce such waves (\S~\ref{sec:sour}). The physics and dynamics of the 
spicules and the detailed evolution of the formation mechanism are 
described in \S~\ref{sec:dyn}. We finish this paper with a 
discussion in \S~\ref{sec:conclusions}. 

\section{Equations and numerical method}
\label{sec:equations}

The numerical model uses the {\it Oslo Stagger Code} (OSC) which 
solves the MHD equations in a domain that stretches from the upper 
convection layer up to the corona. The MHD equations are solved using 
an extended version of the numerical code described in 
\citet{Dorch:1998db,Mackay+Galsgaard2001} and in more detail by 
Nordlund \& Galsgaard at http://www.astro.ku.dk/$\sim$kg and 
\citet{paper1,paper2} (we will refer to the latter two papers as Paper 
I and Paper II, respectively).  The code functions as follows. A sixth 
order accurate method involving the three nearest neighbor points on 
each side is used for determining the spatial partial derivatives combined  
with a fifth order interpolation scheme. The equations are stepped 
forward in time using the explicit third order predictor-corrector 
procedure by \citet{Hyman1979}, modified for variable time steps.  In 
order to suppress numerical noise, high-order artificial diffusion is 
added both in the forms of a viscosity and magnetic diffusivity (Paper 
II).

The radiative flux divergence from the photosphere and lower 
chromosphere is obtained by angle and wavelength integration of the 
transport equation assuming isotropic opacities and emissivities. The 
transport equation assumes opacities in LTE using the solution based 
on four group mean opacities \citep{Nordlund1982}.  The transfer 
equation is formulated for these bins and a group mean source function 
is calculated for each bin. These source functions contain an 
approximate coherent scattering term and an exact contribution from 
thermal emissivity. The resulting 3D scattering problems are solved by 
iteration based on one-ray approximation in the angle integral for the 
mean intensity as developed by \citet{Skartlien2000}. 

In the mid and upper chromosphere OSC includes non-LTE radiative 
losses from hydrogen continua, hydrogen lines and lines from singly 
ionized calcium. These losses are calculated from the local 
net collisional excitation rate multiplied by an escape probability 
that is based on a 1D dynamical 
chromospheric model in which the radiative losses are computed in 
detail \citep{Carlsson:1992kl,Carlsson:1995ai,Carlsson:1997tg,Carlsson:2002wl}. 
In addition to these radiative 
losses in the upper atmosphere we have added an {\it ad hoc} heating 
term to prevent the atmosphere from cooling much below $2000$~K in the 
upper chromosphere. 

For the upper chromosphere and corona we assume optically thin 
radiative losses.  The optically thin radiative loss function is based 
on the coronal approximation and atomic data collected in the HAO 
spectral diagnostics package \citep{HAO_Diaper1994} for the elements 
hydrogen, helium, carbon, oxygen, neon and iron. 

The code includes thermal conduction along the magnetic field. The  
conduction equations are discretized using the Crank-Nicholson method 
and the resulting operator is solved by operator splitting. 
We solve the implicit part of the conduction problem using  
a multi-grid solver. There is also a more detailed description of the 
radiative losses and thermal conduction in Paper I. 

\subsection{Initial and boundary conditions}
\label{sec:condition}

The numerical simulation described here is the same as in Paper II which has a grid 
size of $256\times 128\times 160$ points spanning $16\times 8\times 
16$~Mm$^3$.  The horizontal resolution is uniform at 65~km while the  
vertical grid is non-uniform; thus ensuring that the vertical resolution 
is good enough to resolve the photosphere and the transition region 
with a grid spacing of 32.5~km, while becoming larger at coronal 
heights. At this resolution the simulations have been run for roughly 1.5 
hours of solar time. 

We have seeded the initial model with a magnetic field in which 
sufficient stresses can be built up to maintain coronal temperatures 
in the upper part of the computational domain, as previously shown to 
be feasible by \citet{Gudiksen+Nordlund2004}.  
We introduce a magnetic flux tube into the lower boundary
parallel to the $y$-axis and centered 
roughly at 8~Mm (see right panel of Fig~\ref{fig:distxy}).
This is described in detail in Paper I, \S~3.2.  The 
magnetic flux tube structure is horizontal and axisymmetric:

\begin{eqnarray}
{\mathbf B}_{l}&=& B_o\, \exp \left(-\frac{r^2}{R^2}\right)\, {\mathbf e}_{z}, \label{eq:blong}\\
{\mathbf B}_{t}&=&B_{l}\,r\,q\,{\mathbf e}_{\phi} ,\label{eq:btrans}
\end{eqnarray}

\noindent where $r=\sqrt{(x-x_o)^2+(z-z_o)^2}$ is the radial distance 
to the center of the tube and $R$ is the radius of the tube. ${\mathbf 
  B}_{l}$ and ${\mathbf B}_{t}$ are the longitudinal and transversal 
magnetic fields in cylindrical coordinates, respectively. Note that 
the longitudinal field has a gaussian profile. The 
parameter $q$ is used by \citet{linton1996} and \citet{fan1998} to 
define the twist of the magnetic field. Following \citet{mark2006}, we 
define a non-dimensional twist parameter ($\lambda\equiv q\,R $). 

We have run two different simulations. The 
parameters of the tube, $R$, $B_o$ and $\lambda$, 
for each simulation are given in table~\ref{tab:runs}. The emerging flux tube in simulation
A2 has $B_0=4500$~G at the bottom boundary, with twist 
$\lambda=0.6$. The emerging flux tube in simulation B1 has $B_0=1100$~G and 
no twist. 

As the flux tube enters the computational box, the height of the 
center of the tube ($z_o$) changes in time. At each time step, the 
speed of the flux tube, $({dz_o/dt})$, is set to the average of the 
velocity of plasma inflow at the boundary in the region where the 
magnetic flux tube is located.

We have analyzed simulations A2 and B1 from Paper II searching for 
jet-like features. 
In both simulations, the initial coronal loops are generally 
aligned along the $x$-axis, stretching from magnetic field 
concentrations centered roughly at $x=7$~Mm and $x=13$~Mm (see the 
field line distribution in Fig.~1 in Paper II and right panel of Fig.~\ref{fig:distxy} of  
this paper). Simulation A2 has an 
average unsigned field strength in the photosphere of $16$~G, whereas 
simulation B1 has a field strength of $160$~G. A summary of the simulation 
properties is shown in table \ref{tab:runs}. 

\section{Results}
\label{sec:results}

The simulations produce a variety of different jets which show 
similarities to both type~{\sc i} and type~{\sc ii} spicules. These 
jets are formed by the natural evolution of the 3D model, {\it i.e.,} without imposing any additional 
conditions. Here, we extend the work done by \citet{De-Pontieu:2007cr} 
and \citet{Heggland:2007jt} to 3D and study the effects 
of flux emergence by analyzing two different models, with differing 
values of the pre-existing magnetic field and the magnetic flux tube. We will 
concentrate our attention on the jets that show dynamical behaviour similar to that of 
type~{\sc i} spicules. 
Henceforth we call the jets observed in our 
simulations spicules.

\subsection{Properties}
\label{sec:st}

We found a total of $168$ candidate type~{\sc i} spicules in the two 
different simulations, A2 and B1. To find these features, we use 3D images of 
the transition region and select only jet-like features that reach 
a height larger than $2000$~km above the photosphere. Once 
identified, we plot for each spicule, the temperature as a function of 
time and height (see top panels of Fig.~\ref{fig:traytg}). We then 
fit the trajectory of the transition region (at $T=10^5$~K) with a parabola. We 
reject the jets that do not have a parabolic evolution, or have 
lengths that are less than $500$~km. We are left with $150$  
spicules which we study in more detail below. 

\begin{figure}
   \resizebox{\hsize}{!}{\includegraphics[width=8cm]{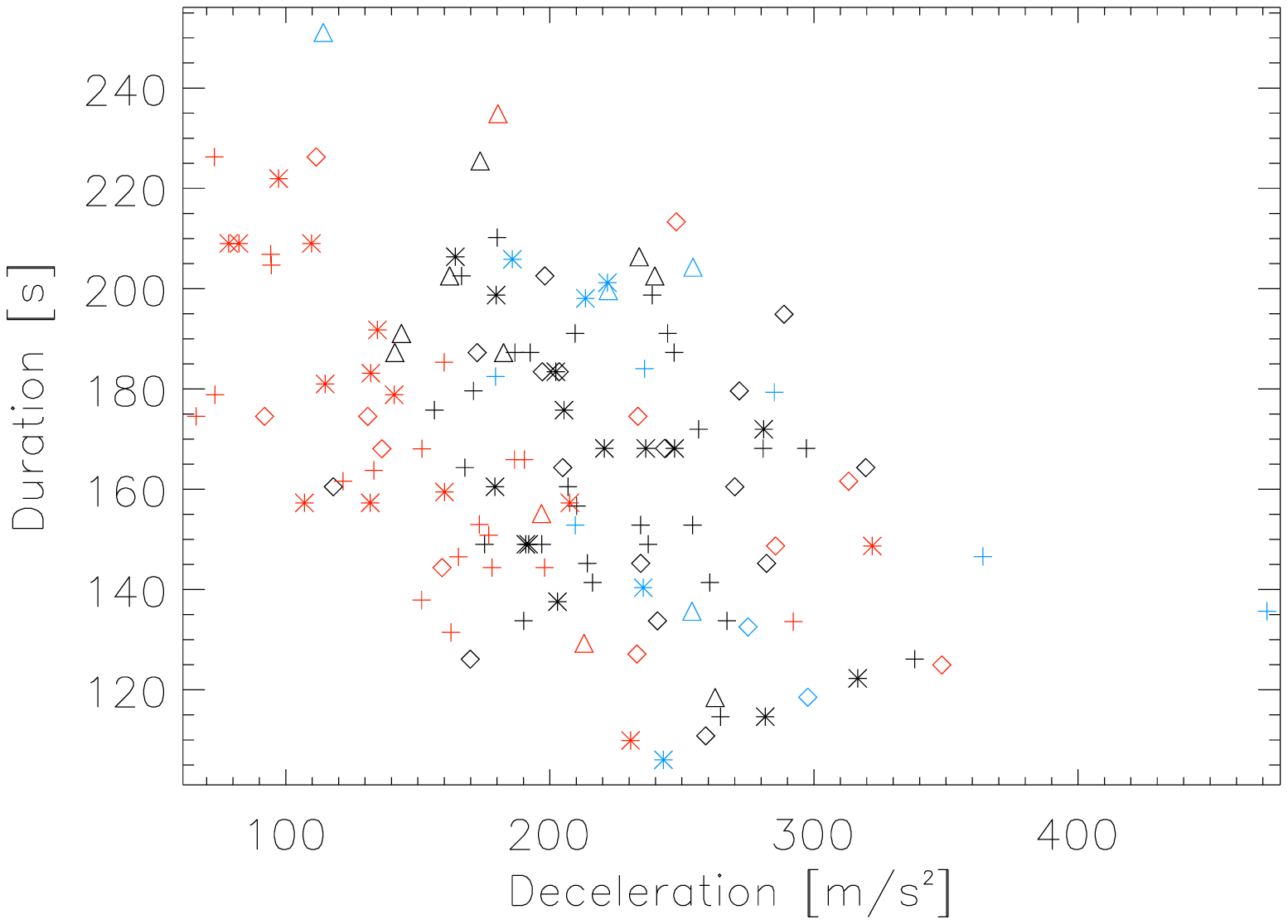}
  \includegraphics[width=8cm]{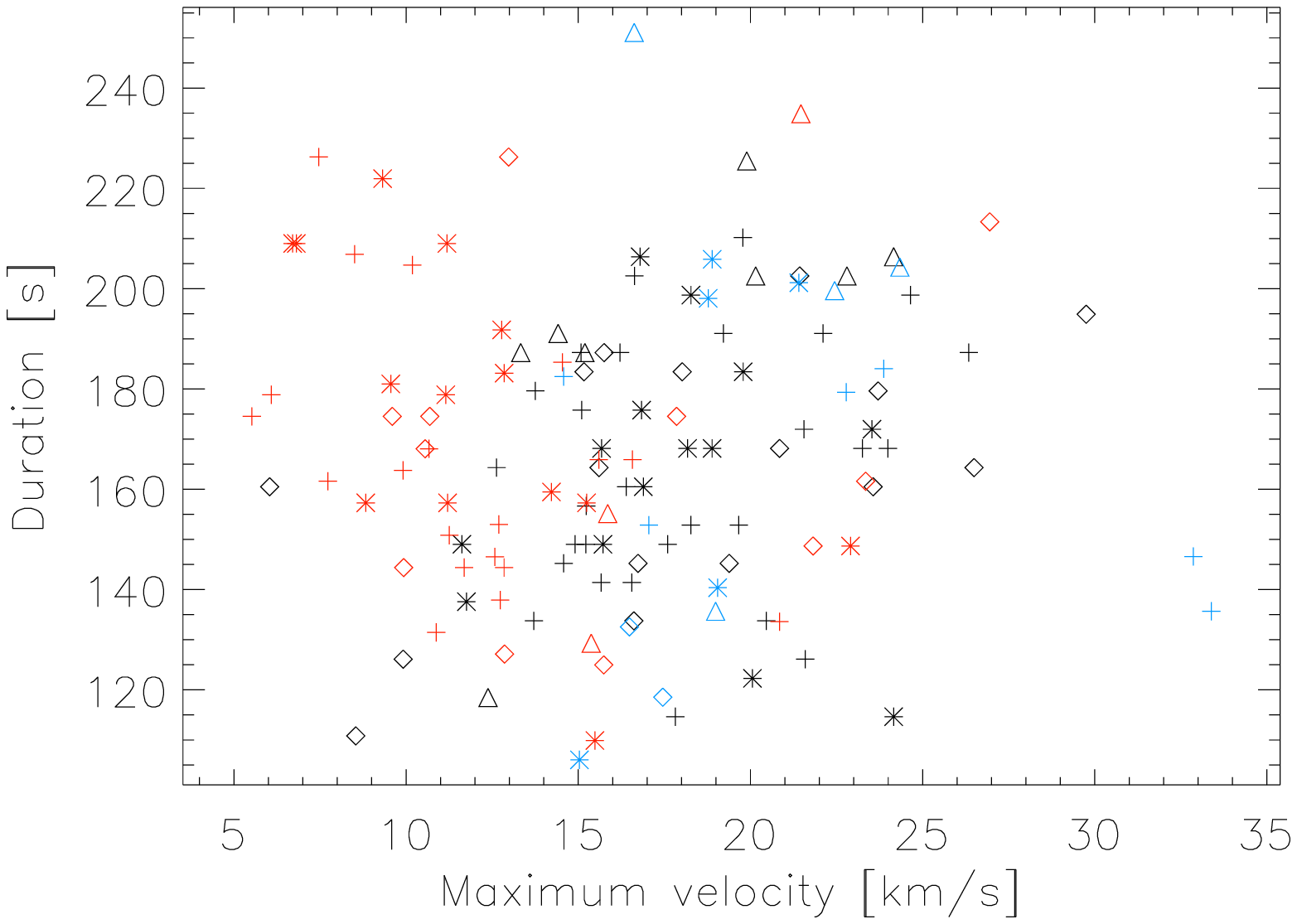}}
   \resizebox{\hsize}{!}{\includegraphics[width=8cm]{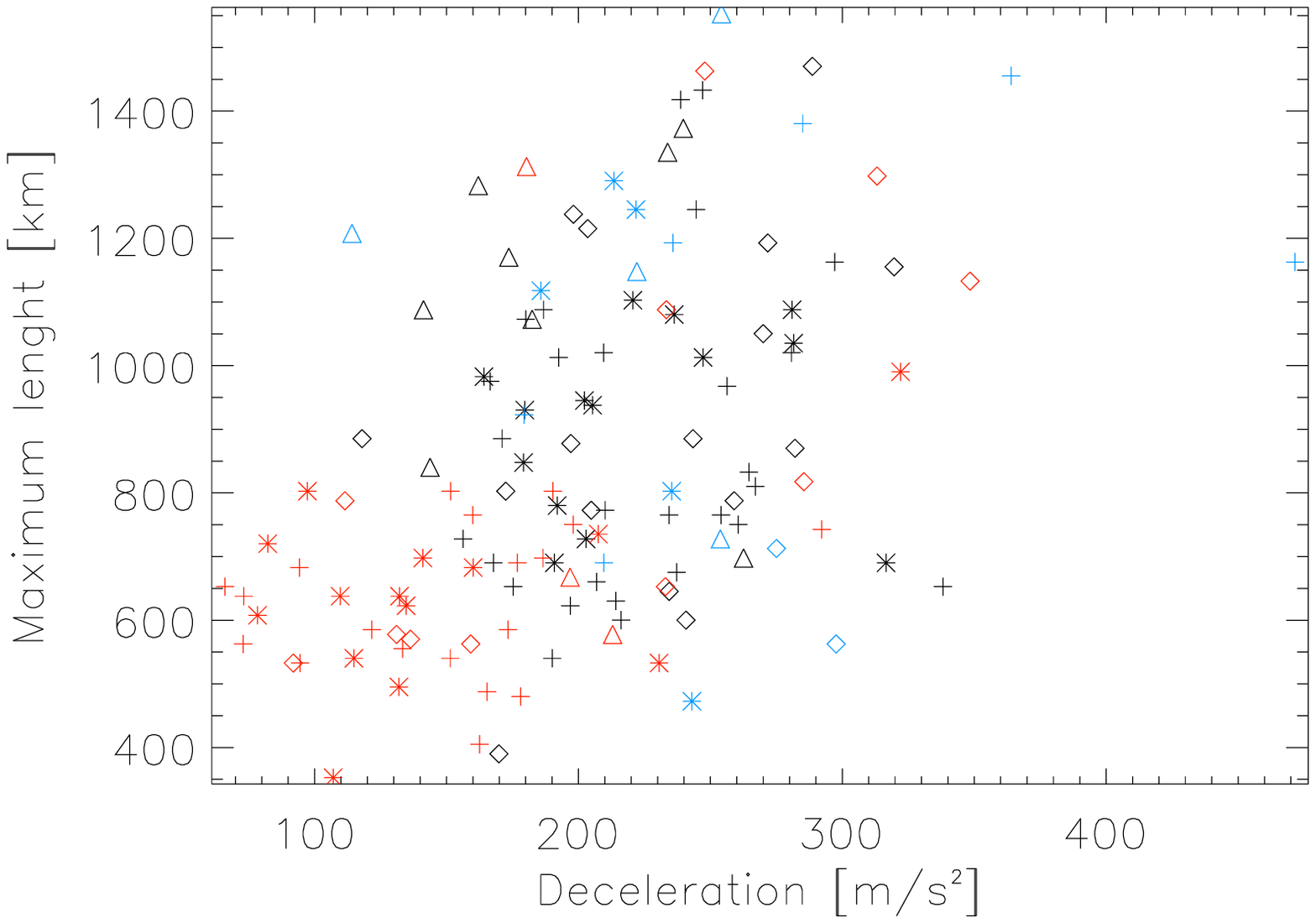}
  \includegraphics[width=8cm]{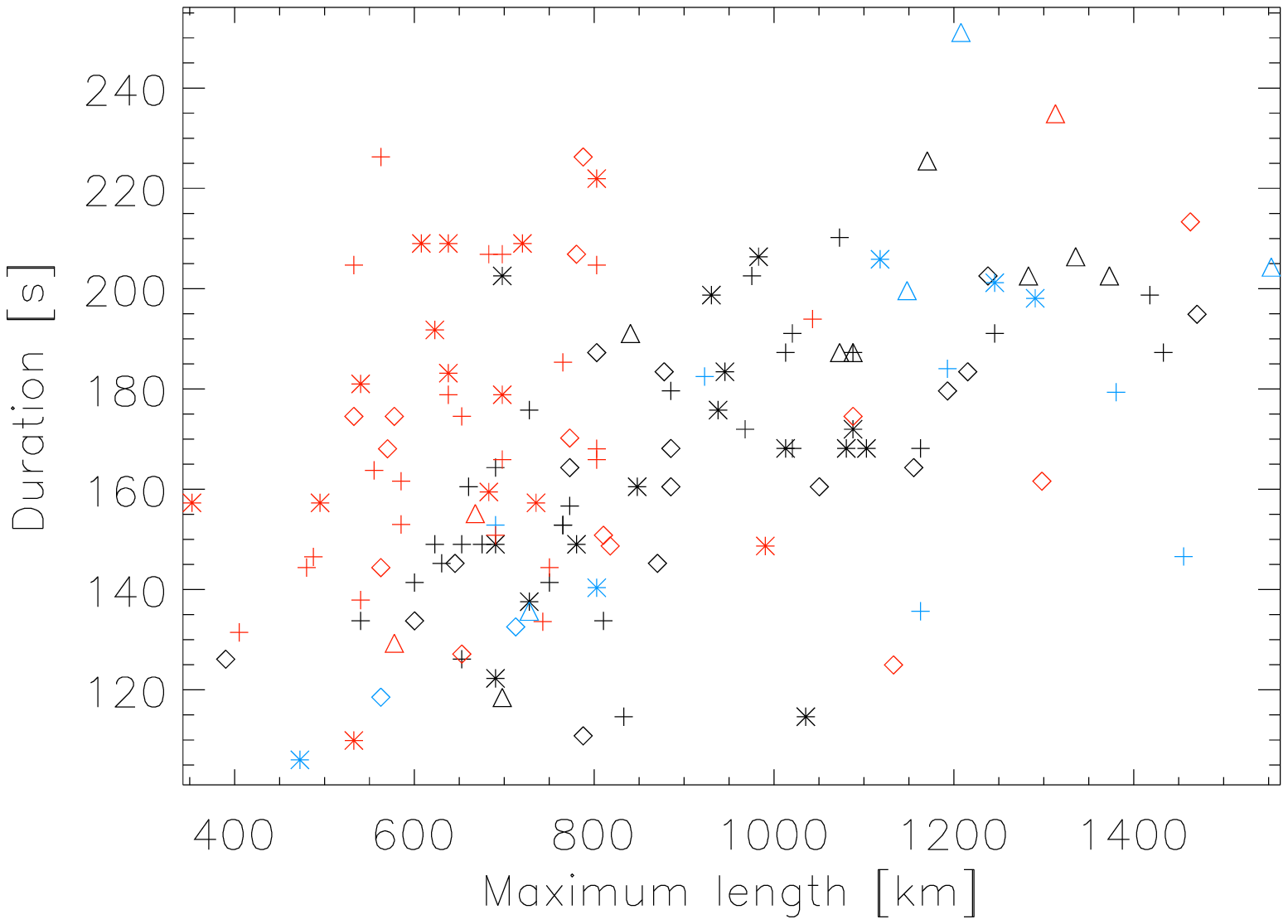}}
   \resizebox{\hsize}{!}{\includegraphics[width=8cm]{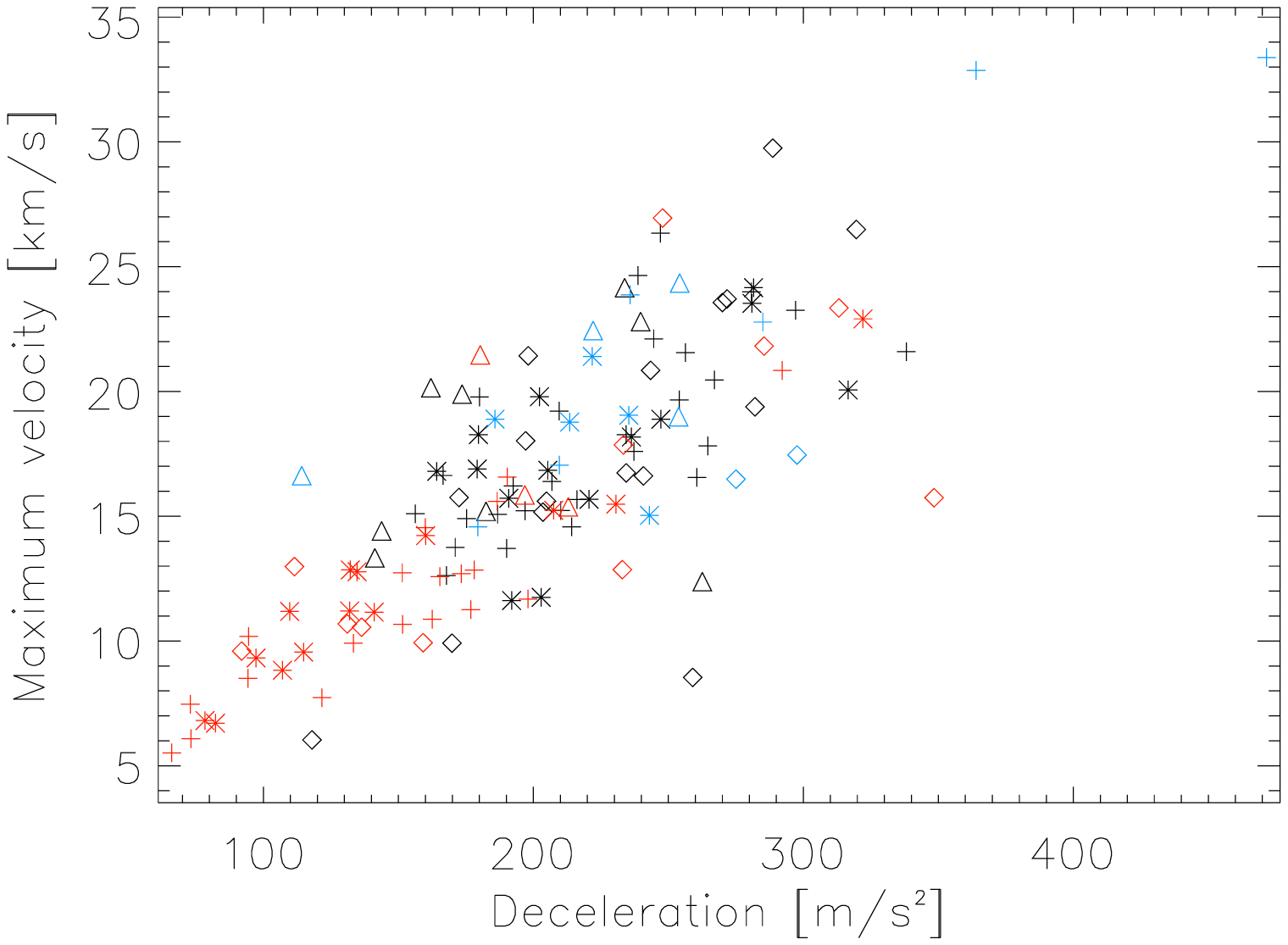}
    \includegraphics[width=8cm]{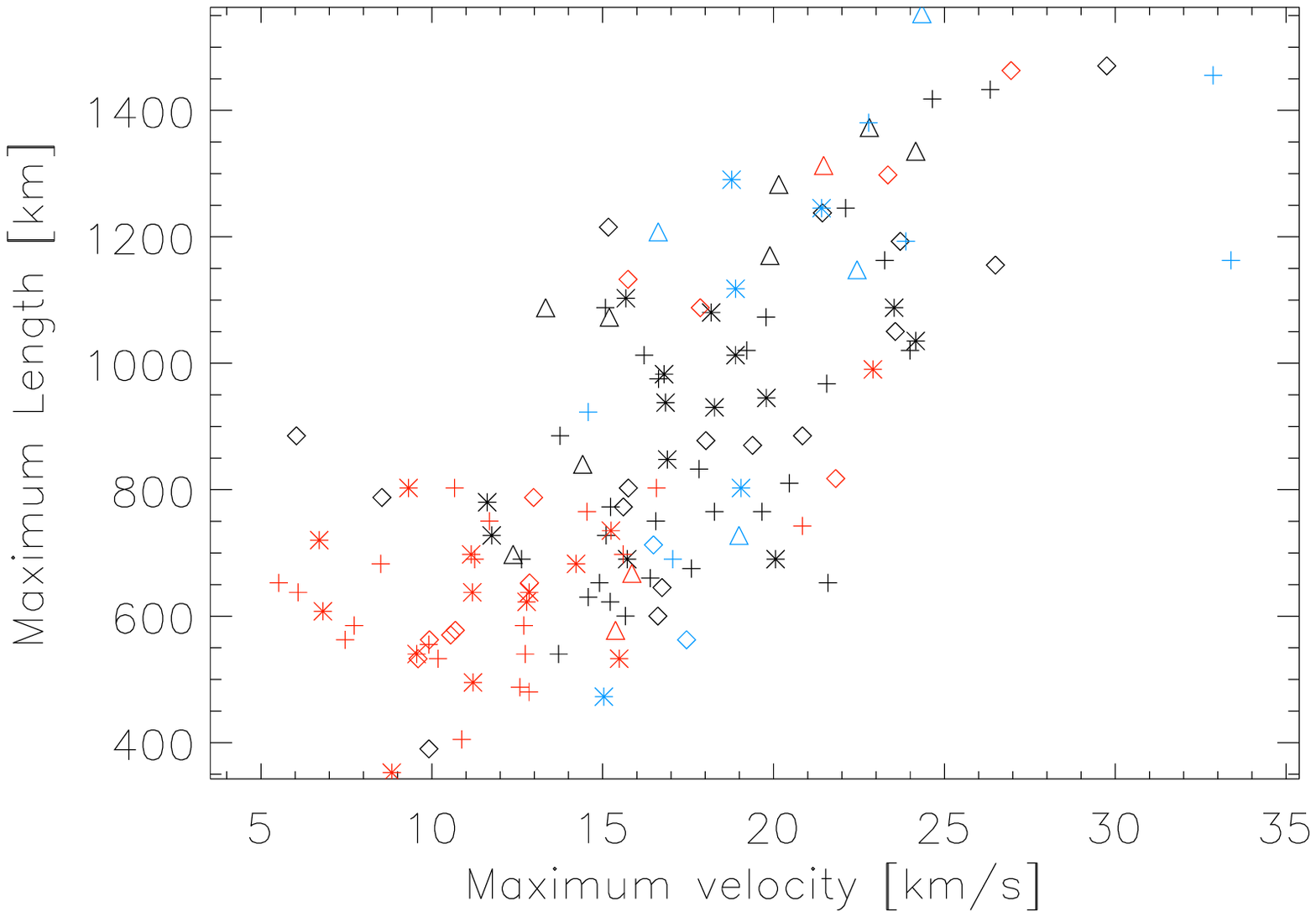}}
  \caption{\label{fig:dist} Scatterplots of parabolic parameters for 
    all 150 type~{\sc i} spicules measured in the simulations. Duration 
    {\it vs.} deceleration of the spicules, maximum length {\it vs.} deceleration, 
    maximum velocity  {\em vs.} deceleration, duration {\it vs.} maximum 
    velocity, duration  {\it vs.} maximum length and 
    length  {\it vs.} maximum velocity are illustrated from top to bottom 
    and left to right. The spicules from simulation A2 during the temporal range 
    $[300-3200]$~s are shown in black and the spicules during the 
    temporal range $[3800-5100]$~s in blue. Spicules from simulation 
    B1 are shown in red. The deduced driving mechanism  
    is shown with symbols: collapsing 
    granules (plus sign), photospheric magnetic energy release 
    (rhombus), chromospheric magnetic energy release (triangle) 
    and other or unidentified mechanisms (asterisk).} 
\end{figure}

Once the motion of the transition region in these spicules has been fit 
with a parabolic trajectory, we measure the following parameters: maximum length, 
duration, maximum velocity, deceleration and the 
angle of the magnetic field with respect to the 
vertical. Given that most spicules are close to vertical and quite wide, we
take into account their non-verticality by correcting these
parameters with the cosine of the magnetic field angle with respect to the vertical.
Figure~\ref{fig:dist} shows scatterplots illustrating the 
relation between some of these parameters; for instance, deceleration 
{\it vs.} duration, deceleration {\it vs.} maximum velocity of the spicule etc. 

Comparison of Fig.~\ref{fig:dist} with Figs.~12 and 13 from 
\citet{De-Pontieu:2007cr} shows remarkable similarities of the 
correlations between the various parameters of the jets in 
our simulations and those found for dynamic fibrils from 
observations. Similar to the dynamic fibrils, we find that the spicules in 
the simulations with longer duration have smaller deceleration. In 
addition, spicules that are longer have higher deceleration, and the 
higher the deceleration, the higher the maximum velocity is. Spicules 
with longer duration are typically longer, and the longer spicules 
have higher maximum velocity. There does not appear to be 
much correlation between duration and maximum velocity. All of 
these relations fit well with those observed in dynamic fibrils, quiet 
Sun mottles and type~{\sc i} spicules. Moreover, the range of values  
for deceleration and maximum velocities are 
similar for observations and simulations.  

There are also differences 
between the simulations and observations, particularly with respect to 
the range of values that are found for duration and length. These are 
found to have a smaller range in the simulations than in the 
observations. Most durations are around 3 minutes, instead of the wide 
range of durations between 2 and 8 minutes found in the observations.  
All of the spicules 
that form in the models are broader than in the observations, most 
likely because the numerical resolution is relatively low. The 
simulated spicules are also shorter in length than in the 
observations. These discrepancies are discussed in \S~\ref{sec:conclusions}. 

The role of the magnetic configuration and/or field strength is 
illustrated by the fact that the values we find in the simulations 
seem more in line with those found in region 2 of 
\citet{De-Pontieu:2007cr}, {\it i.e.}, a denser plage region,  
which showed shorter fibrils with durations of order 
180 s. In addition, we find that the maximum velocity, deceleration, 
and maximum length are all smaller for the spicules we find in 
simulation B1 (red symbols in 
Fig.~\ref{fig:dist}). We find that the number of spicules per 
unit time is higher with higher ambient magnetic field ({\it i.e.}, simulation 
B1). There seems to be a 
trend towards a slightly different correlation between duration and 
maximum length, and duration and deceleration for the two different 
runs. We generally do not see any significant differences in these 
correlations for the two different periods, during and after flux 
emerges through the photosphere, in simulation A2 (black 
and blue in Fig.~\ref{fig:dist}, respectively). 

While all of these spicules are driven by magnetoacoustic shocks, the 
waves that develop into these shocks are caused by a variety of 
drivers.  We describe in detail in \S~\ref{sec:sour} the different 
drivers we have identified from the simulations: collapsing granules 
(plus signs in Fig.~\ref{fig:dist}), magnetic energy 
release in the photosphere (rhombi in Fig.~\ref{fig:dist}) 
and magnetic energy release in lower chromosphere  
(triangles in Fig.~\ref{fig:dist}).  
There 
seem to be some differences in correlations for spicules caused by 
different drivers. The spicules driven by 
chromospheric magnetic energy release (triangles) are located in the 
upper part of the scatter plot showing the correlation between 
deceleration and maximum velocity (middle left panel of 
Fig.~\ref{fig:dist}) and of the scatterplot showing the correlation 
between maximum velocity and maximum length (bottom right panel of 
Fig.~\ref{fig:dist}). Other than these differences, the various 
driving mechanisms do not seem to lead to any other differences in the 
correlation plots. 

\begin{figure}
 \resizebox{\hsize}{!}{\includegraphics[width=8cm]{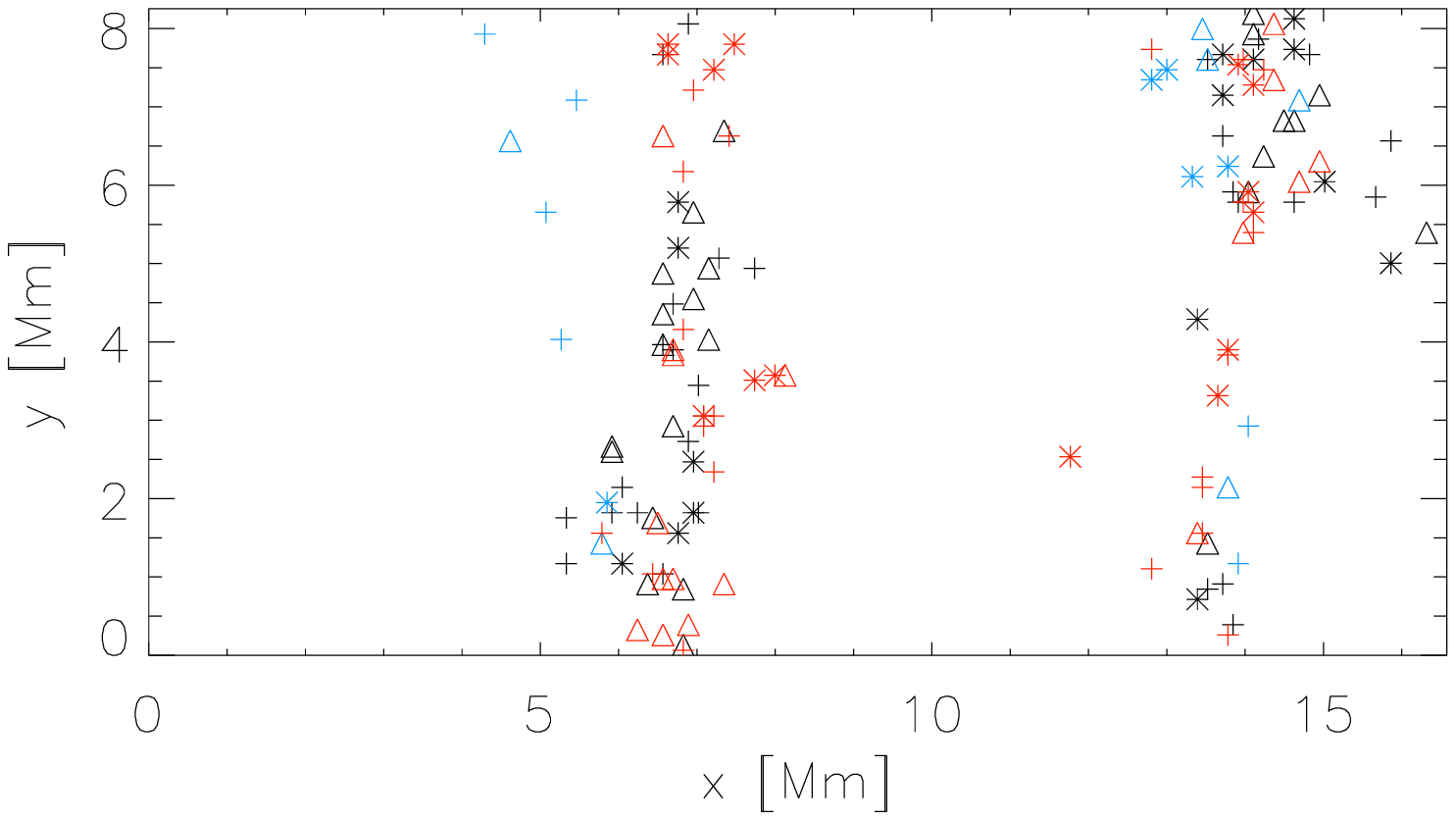} 
 	 \includegraphics[width=8cm]{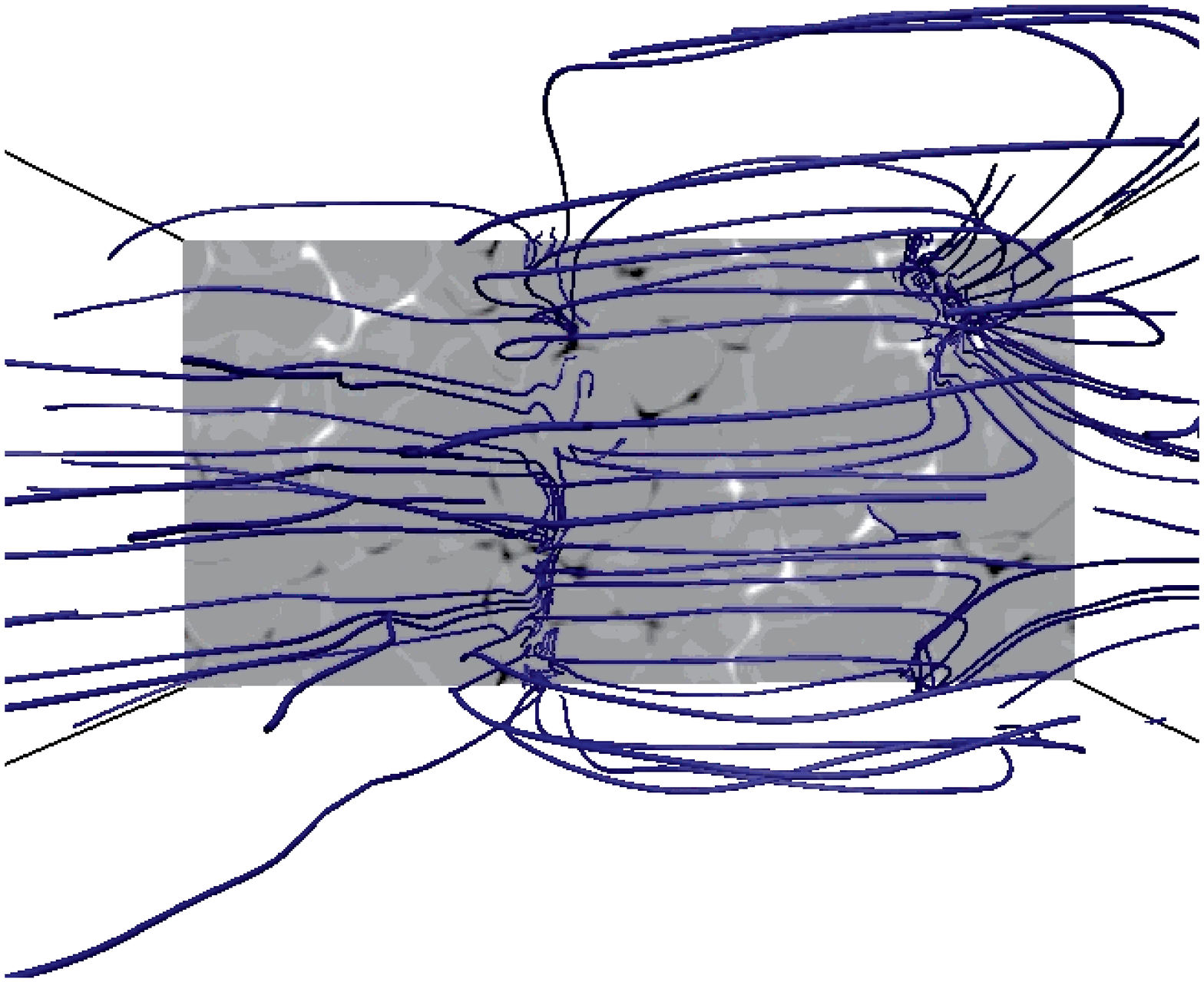}}
 \caption{\label{fig:distxy} Map showing the ($x,y$) locations 
 of the spicules (left panel). The spicules from simulation 
 A2 during the temporal range [$300-3200$]~s are shown 
 in black and the spicules during the temporal range 
 [$3800-5100$]~s in blue. Spicules from  
 simulation B1 are shown in red.  
 The driving mechanism is shown with symbols: collapsing granules (plus symbol),  
 magnetic energy release (triangle) and other mechanisms (asterisk). The right panel 
 shows a view from above of the computational domain of the simulation A2
 showing field lines traced from the corona (blue lines) and the vertical magnetic 
 field strength in the photosphere (grey-scale).} 
\end{figure}

When considering the location of the occurrence of spicules we find that all 
the spicules we have identified are concentrated at the foot  
points of coronal loops as seen in Figure~\ref{fig:distxy}. In the right panel of 
Figure~\ref{fig:distxy} we draw a set of field lines computed by integrating from  
randomly chosen points in a horizontal plane high in the corona. These field lines all 
tend to have their foot points along two lines oriented roughly in the  
$y$-direction. In addition, the field lines are roughly vertical as they pierce the  
photosphere. It is along these field lines that our spicules propagate. 
During the evolution of the flux emergence in simulation A2 the left foopoints of 
the coronal loops move leftward. This is clearly reflected in the spicules 
distribution (Fig.~\ref{fig:distxy}): spicules after flux emergence, shown in blue, are 
further left than the spicules that occur before and during (black). 

\subsection{Driving mechanism}
\label{sec:sour}

\begin{figure}
 \resizebox{\hsize}{!}{\includegraphics[width=8cm]{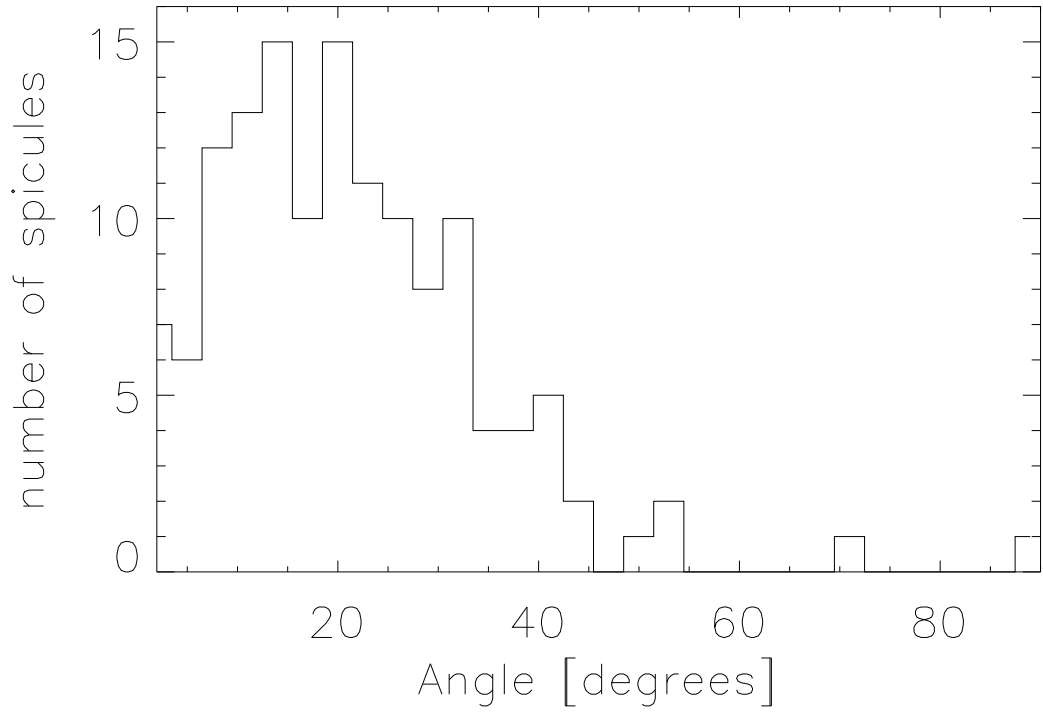}}
 \caption{\label{fig:tan} Histogram of the angle between the vertical axis and the 
   direction of the field lines which go through the spicules. 
   Most of the spicule angles are between $10^o$  
   and $30^o$.} 
\end{figure}

Analysis of our simulations tells us that the spicule-like 
features are produced by magnetoacoustic shocks which form when waves 
that are generated in the photosphere or lower chromosphere propagate 
through to the chromosphere and shock because of the rapid 
drop in density with height. The magnetoacoustic shock drives plasma 
upwards, leading to a rapid upward motion of the transition region, 
followed by a downward motion after passage of the shock (see 
\S~\ref{sec:prof} for details). This behavior is identical to the 
scenario outlined by \citet{Hansteen+DePontieu2006}. The time delay 
between the generation of the wave and the formation of the spicule 
depends strongly on the length of the trajectory. For instance, a wave 
generated in the photosphere takes roughly 3 minutes to reach the 
transition region, whereas it only takes roughly 1.5 minutes from the 
lower chromosphere. It seems that two conditions must be satisfied  
in order to produce 
spicule-like features in our simulations: 1) The waves require strong  
magnetic flux concentrations to guide them upward; 2)  
The field must extend into the corona which in our models occurs only for 
field lines with angles with respect to the vertical between $0-30^o$.  
Figure~\ref{fig:tan} 
illustrates the range of angles from the vertical in all 
spicules we analyzed. We find that the angles do not depend 
on the magnetic field strength and/or flux emergence properties. 
The mostly vertical field we measure might be caused by a selection  
effect since the selection criterion 
we use to identify spicules requires a minimum height of 2000~km. 
Given the short lengths of the spicules produced, this criterion favors more 
vertically oriented spicules, since these are the only field lines that reach the  
corona (see also \S~\ref{sec:conclusions}).  

In order to determine the driver, we analyze one or more magnetic field 
lines that follow the spicule's long axis, and we identify the shock 
front. We trace the shock front back until the time when the shock is 
formed. Once we localize the place and instant of shock formation, we 
analyze the surroundings of that region and try to find which kind of 
process perturbs the region enough to produce a wave. In most cases 
this is quite difficult because of the complex mix of events that 
occur in our 3D simulations. In many cases we find that the wave could 
be excited by several different physical processes, whereas in other cases we 
are unable to identify a dominant cause; we will refer to these unidentified drivers 
as  ``other mechanisms''. Generally, we find that the 
most frequent processes that cause these waves are: p-mode 
oscillations, collapsing granules, breaking granules, flux emergence 
through the photosphere, magnetic energy release in the photosphere or 
in the lower chromosphere. This is by no means a complete list, because, 
as just mentioned, it is often difficult to identify the driver. 

The easiest 
drivers to identify are collapsing granules, magnetic energy release
in the photosphere, and magnetic energy release in the 
chromosphere. However, in most cases, it is difficult to have 100\% 
confidence in the identification. The number of spicules caused by the 
different drivers is different for both simulations: for simulation A2, the 
percentage of spicules caused by collapsing granules, magnetic energy 
release in the photosphere, magnetic energy release in the 
chromosphere and other mechanisms is respectively, 26\%, 29\%, 19\% 
and 26\%. For simulation B1 with higher ambient field, these numbers are 42\%, 
26\%, 6\% and 26\%. Magnetic energy release in the chromosphere thus 
contributes significantly less to the formation of spicules when the 
ambient magnetic field is stronger (perhaps because most magnetic 
energy release occurs lower down, since emerging flux encounters 
significant flux at lower heights). In both simulations a significant amount 
of flux emerges. This contributes in different ways to the excitation 
of the shock waves that drive the spicules. There is a peak in the number 
of spicules driven by collapsing granules when the flux is first 
entering the photosphere from below. Afterwards, spicules driven by 
magnetic energy release in the photosphere dominates, since there 
appears to be an interaction between the rising flux tube and the 
ambient field. Generally, the spicules driven by collapsing granules 
seem more prevalent in the weak ambient field simulation (A2) where the flux 
tube perturbs the granulation pattern more profoundly. 

\begin{figure}
	\includegraphics[width=8cm]{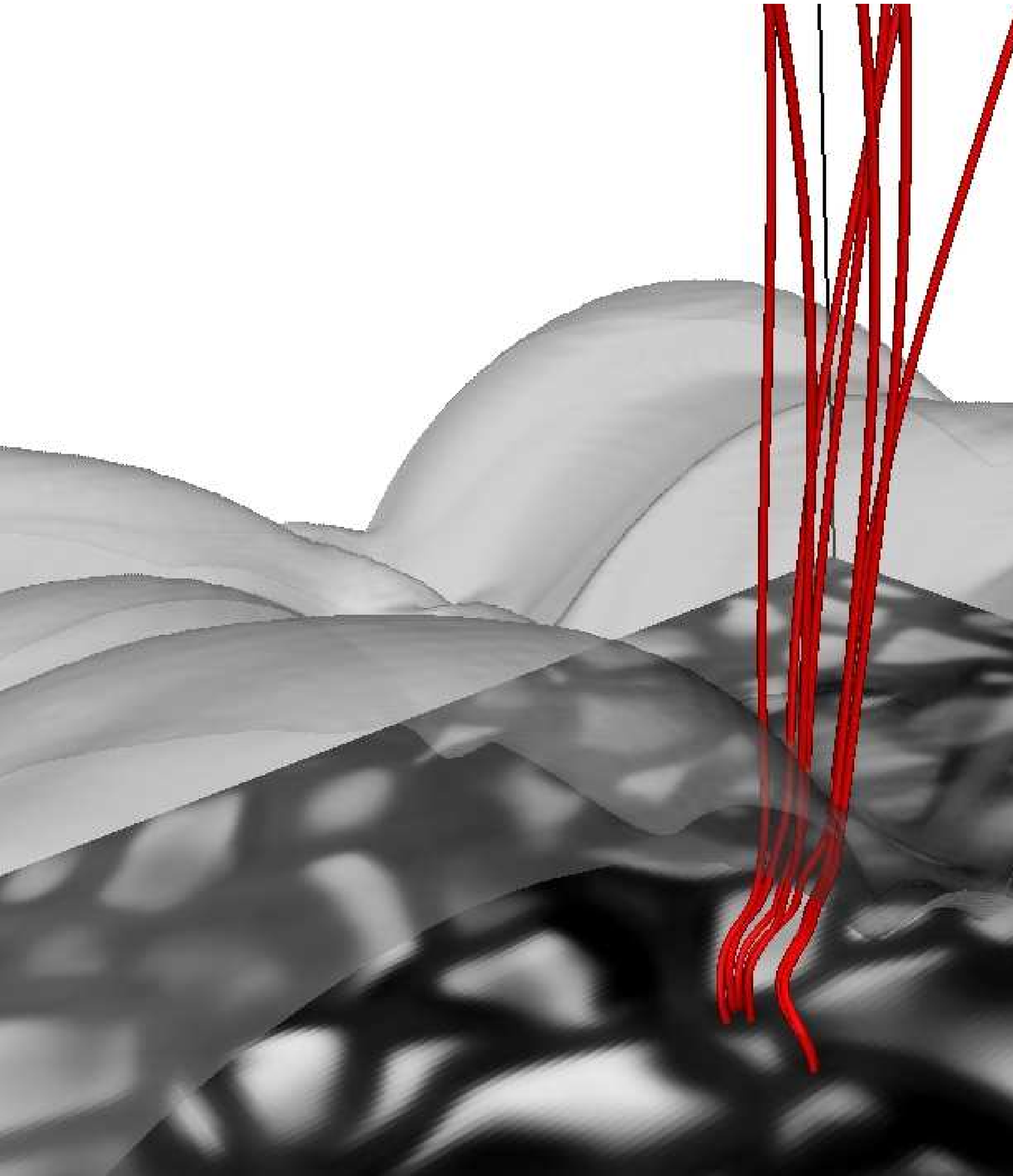}
		\includegraphics[width=8cm]{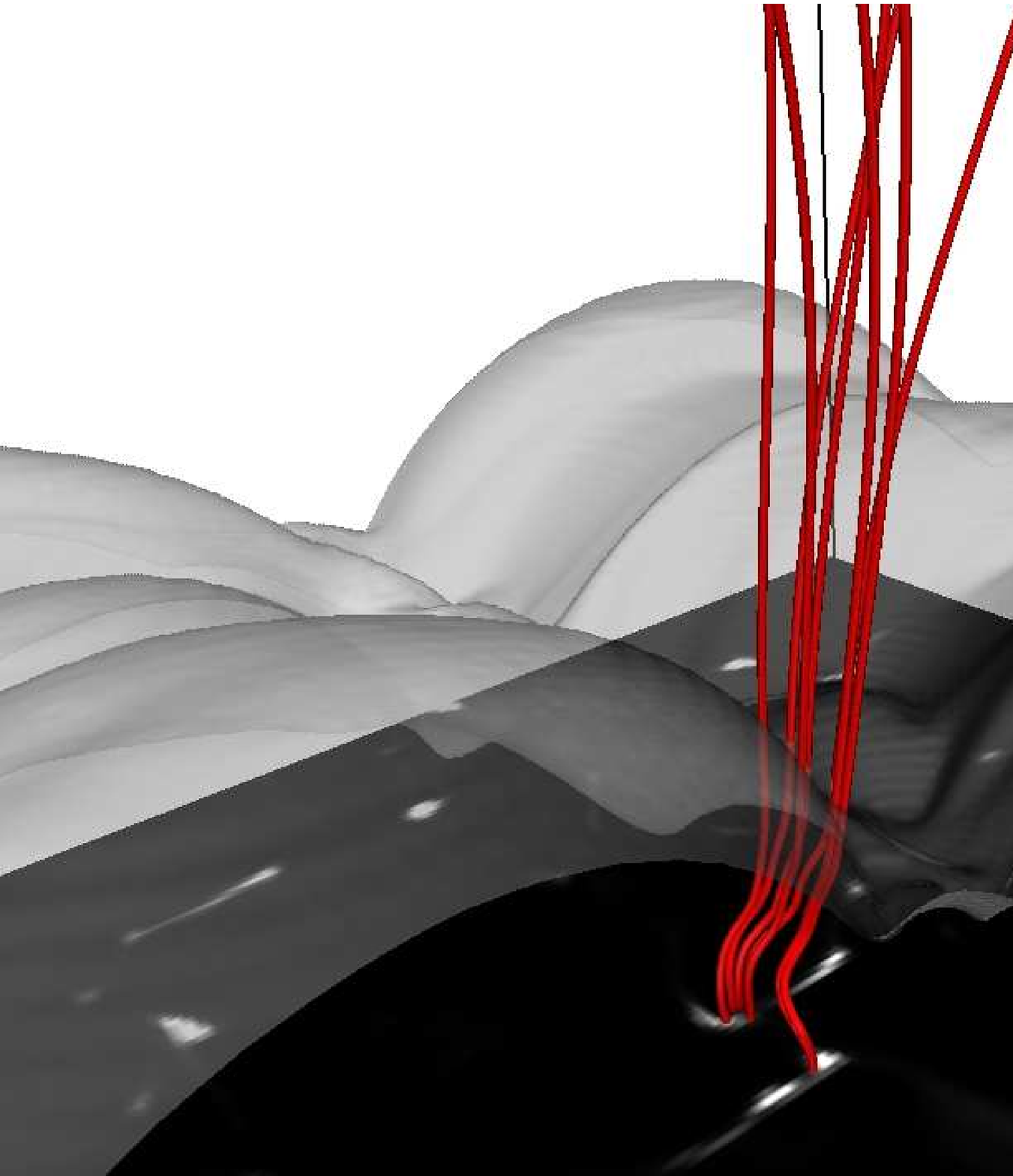}
                \caption{\label{fig:collspic} Magnetic field geometry and connection
                 to the photosphere of a spicule driven by a collapsing granule. Temperature in the 
                  photosphere is shown in a flat surface in grey color scale in the left 
                  3D image. In the right 3D image, the absolute magnetic field strength in the 
                  photosphere is shown with a flat surface in grey color scale with a range
                  	of [$0-12$]~G. The 
                  (semitransparent) grey isosurface is the transition region at 
                  $T=10^5$~K. The red lines come from a spicule which 
                  is formed by a collapsing granule which is 
                  surrounded by the field lines. The spicule forms 
                  along the strongest concentration of flux and where 
                  the field lines go into the corona.} 
\end{figure}

A collapsing granule produces a perturbation or  
wave that affects temperature and pressure 
and propagates in all directions (see \S~\ref{sec:prof}) \citep{Skartlien:2000lr}. It can 
perturb one or more flux concentrations and  
produces spicules when the disturbance propagates along flux concentrations. In 
figure~\ref{fig:collspic} we show an example where a spicule is formed 
because of a wave triggered by a collapsing granule that is surrounded 
by two different flux concentrations in the photosphere (right 
panel). It is unclear which of the two flux concentrations is dominant 
in the spicule formation because the numerical resolution is too 
low. We see that the field lines along which the spicule forms are 
quite vertical and go into the corona. The presence of a strong flux concentration and  
field lines that enter into the corona, generally allow for easier 
propagation of the perturbations from the photosphere into the 
chromosphere and the corona, thus facilitating the formation  
of spicules in our simulations.  

The deposition of thermal energy as a result of magnetic energy 
dissipation can produce a perturbation or wave that can steepen into a 
shock when propagating into the chromosphere along a large flux 
concentration (see \S~\ref{sec:prof}). We tentatively call these 
events ``magnetic energy release''. They usually are not as 
concentrated in space and time and thus as explosive as typical 
reconnection events observed in the solar atmosphere ({\it e.g.}, explosive 
events in which the dominant component of the field shows opposite 
polarity). In fact, in many cases we do not find opposite polarity 
field lines at the location of the driver. We often find more gentle 
dissipation of currents to be the driving mechanism. The magnetic 
energy release sources that we manage to identify are concentrated 
in the upper photosphere and lower chromosphere. In what follows 
we call magnetic dissipation related events ``photospheric magnetic 
energy release'' if the event occurs at heights of $0<z<500$~km, and 
``chromospheric magnetic energy release'' for $500<z<1000$~km. 

To study where current dissipation or magnetic energy release takes 
place in our simulations, we use the following parameter: 

\begin{eqnarray}
R_d=\frac{|{\mathbf J}|}{\sqrt{P}} \label{eq:rd}
\end{eqnarray}

\noindent where $P$ and ${\mathbf J}$ are the gas pressure and the current density, 
respectively.
This parameter shows the discontinuity of the
magnetic field since it is proportional to the magnetic field
strength (and inversely proportional to the length scale over which
it changes). It is a good proxy to identify where the current is
important and the location of current sheets. 
Usually, the discontinuity of the magnetic field is shown with $|{\mathbf J}|/|{\mathbf B}|$. However, 
several regions in our models have field that is weak or
almost zero, {\it e.g.}, in the upper part of the corona or in the
convection zone, which makes it difficult to evaluate $|{\mathbf J}|/|{\mathbf B}|$ numerically. 
This is why we normalized with the square root of the pressure instead 
of using the magnetic field strength.

\begin{figure}
	\includegraphics[width=8cm]{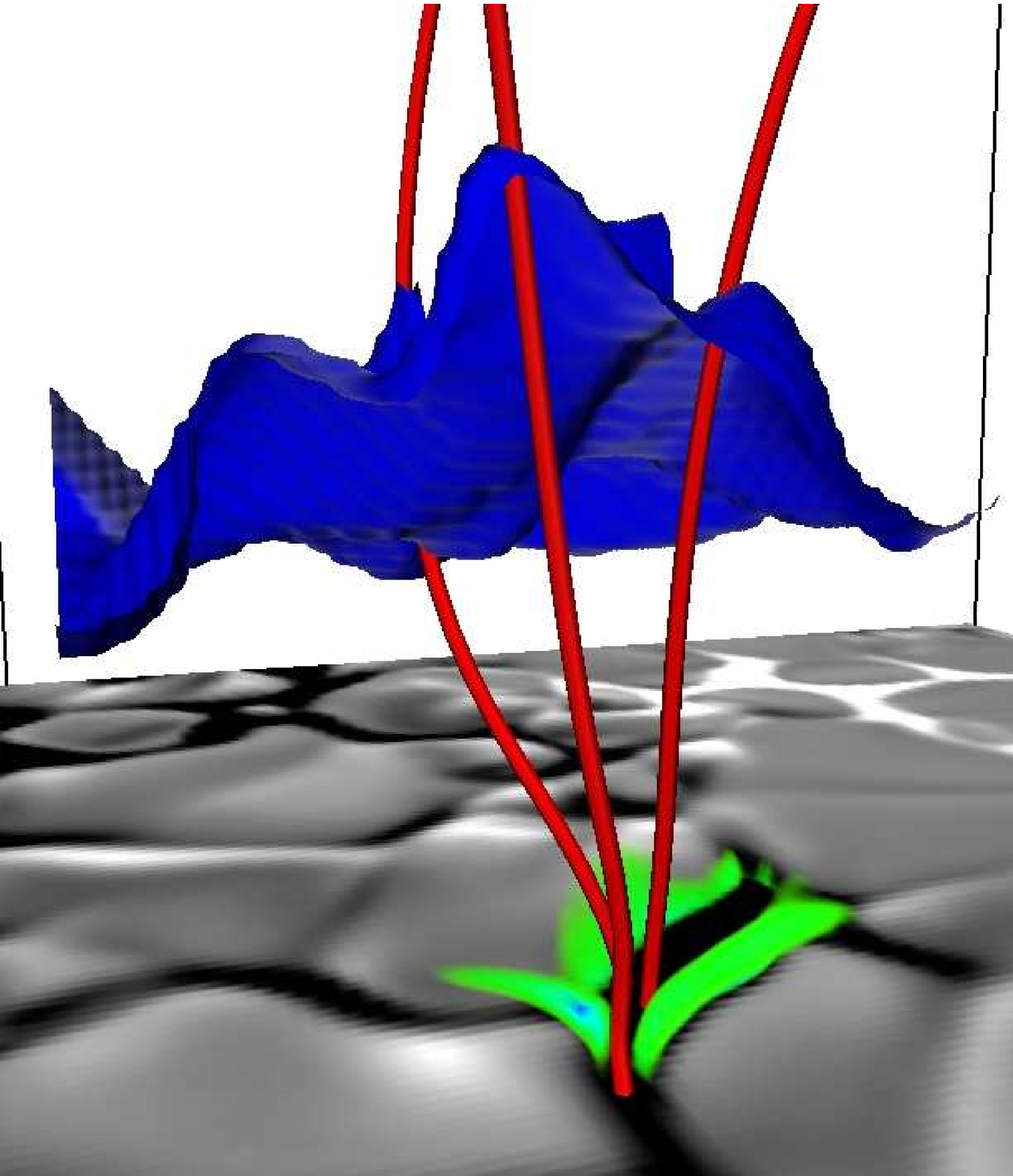} 
	\includegraphics[width=8cm]{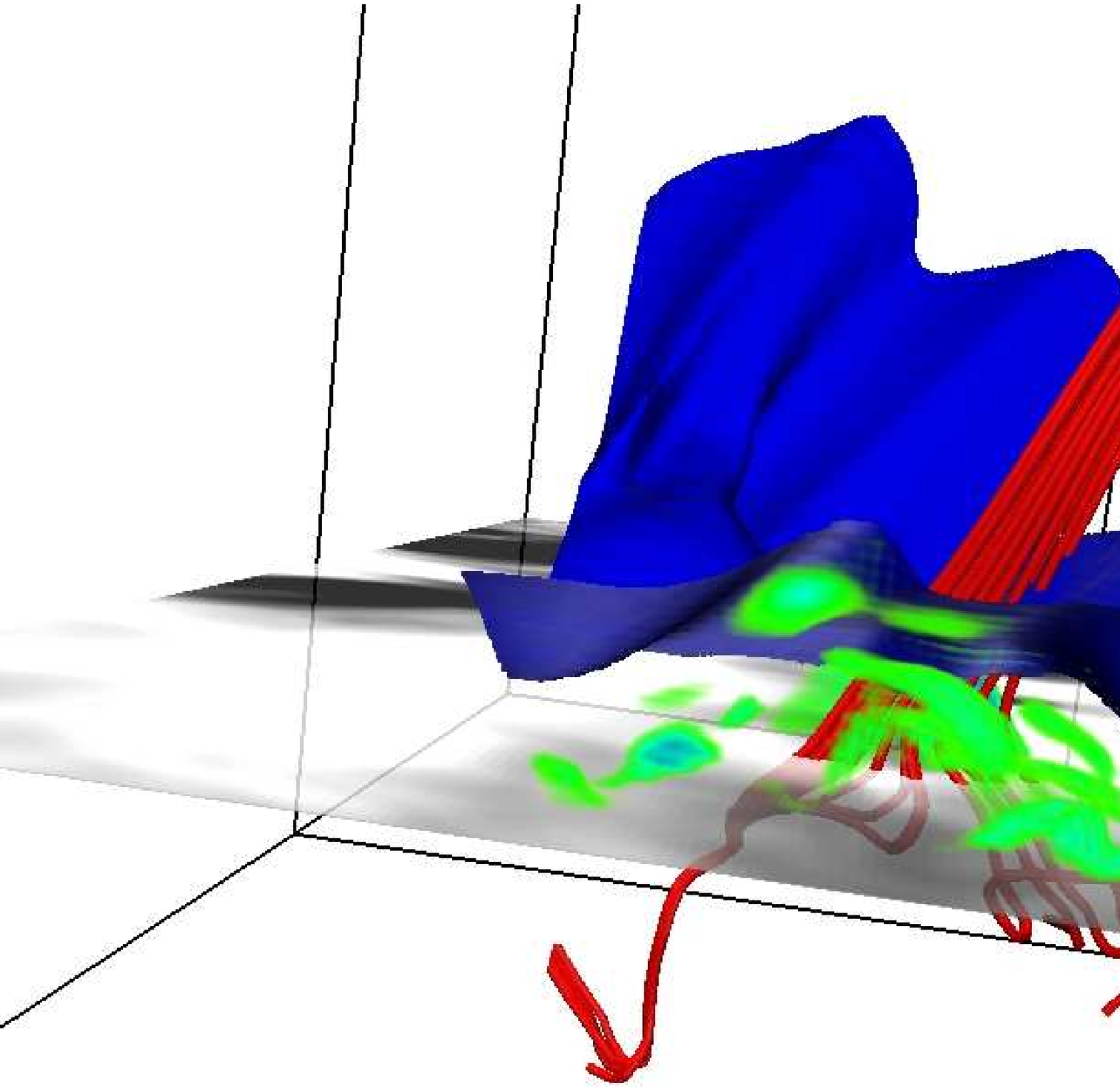} 	
        \caption{\label{fig:recspic} Magnetic field geometry and connection to the 
        photosphere of three different spicules driven by magnetic energy release 
        in the photosphere (left panel) and magnetic field geometry and connection to the 
        chromosphere of a spicule driven by magnetic energy release 
        in the chromosphere (right panel). Vertical magnetic field strength 
          is shown in grey color scale; in the left 3D image at the 
          photospheric level with a range of $\pm 12$~G, and in the 
          right 3D image at the chromosphere level ($z=1$~Mm) with a 
          range of $\pm 5.6$~G. The transition region ($T=10^5$~K) 
          is shown with the blue isosurface. The current over the 
          square root of the gas pressure ($R_d$, see Eq.~\ref{eq:rd}) 
          shows regions where magnetic energy dissipation is high (green  
          color). The three field 
          lines shown in red in the left image end up in three 
          different spicules but trace back to the same driver in the 
          photosphere, a magnetic energy release event. The lines are 
          concentrated in the intergranular lane in association with a 
          strong magnetic flux concentration. The right image shows an 
          example of a spicule driven by magnetic energy release in the 
          chromosphere. Here the field lines are concentrated in a 
          flux concentration at chromospheric heights, but spread out 
          significantly at lower heights.} 
\end{figure}

Figure~\ref{fig:recspic} shows two examples in which magnetic energy 
is released and spicules are formed as a result. In the left image, 
three different spicules occur around the same time. They can all be 
traced back to a magnetic event that takes place in the photosphere, 
in a single flux concentration. The right panel shows the formation of 
a spicule as a result of magnetic energy release in the lower 
chromosphere (green-blue color scale, where $R_d$ is high). 
All field lines that 
are involved in the spicules are associated with locations of high 
$R_d$, {\it i.e.}, the lines are associated with magnetic energy release events. In 
the example on the left, we see that the same magnetic energy release 
process (in the photosphere) can trigger more than one spicule at the 
same time. When the magnetic energy release occurs higher up (right 
image), it seems more difficult to produce more than one spicule from 
the same event. This may be because the path from the source to the 
transition region is shorter and does not allow differential 
propagation of the wave that forms the spicule(s).  

We observed of the order of 20 examples where one specific excitation event seems to 
produce several spicules. The source (collapsing granule, p-modes, and 
magnetic energy release in the photosphere) is able to perturb several 
magnetic flux concentrations (or even the same flux concentration) and 
produces several spicules as a result. These spicules are excited by 
the same driver and occur close in time and space. 
This phenomenon is similar to what has 
been reported in observations of dynamic 
fibrils by \citet{De-Pontieu:2007cr}. 
  
Returning to the location of spicule generation shown in figure~\ref{fig:distxy}, 
we note that during flux emergence through the photosphere 
it is more common to find spicules driven by collapsing granules 
at the footpoints of the loop situated within $x=[5-8]$~Mm, where the 
emerging flux is located. This is presumably because the granular dynamics are 
heavily perturbed by the flux emergence. At later times (locations plotted 
in blue) flux emergence does not perturb the granulation pattern as much, 
and the occurrence and distribution of the spicule types is similar in 
both of the footpoint bands located at $x=[5-8]$~Mm and at $x=[13-15]$~Mm  
in the photosphere. 
In simulation B1, the strength of the emerging magnetic flux is lower and the 
ambient magnetic field is stronger than in simulation A2. Thus, the 
granulation pattern does not change as much with flux emergence, 
and the distribution of spicules driven by collapsing granules is similar in  
both bands of loop footpoints for simulation B1. In contrast, the spicules 
driven by collapsing granules in simulation A2 are more frequently located in 
the region $x=[5-8]$~Mm. This is because in this simulation the flux emergence close to that
region impacts the granulation more significantly. 

The spicules driven by magnetic energy  release occur more frequently in the left band of footpoints. This is true
in both simulations and could be a result of the interaction between the rising magnetic flux and the 
ambient field. 

\subsection{Dynamics}
\label{sec:dyn}

In general, all the spicules studied in this paper are the result 
of upwardly propagating shock waves. In order to understand the dynamics 
of spicule evolution in more detail, and in particular, how 
spicules are excited by the different drivers, we use 
figures~\ref{fig:traytg}-\ref{fig:traycr}. We will first explain the terms shown in the 
figures, which then will be used to study spicule evolution and 
analyze how the various drivers cause spicules. 

Figure~\ref{fig:traytg} shows three clear examples of spicule-like 
jets driven by a collapsing granule (first column), by magnetic energy 
release in the photosphere (second column), and by magnetic energy 
release in the chromosphere (third column). From the top to bottom row 
we show the logarithmic temperature, upflow velocity and vertical 
acceleration as a function of height and time. 
The particle trajectory (blue line in the first row of Fig.~\ref{fig:traytg}) is 
derived by calculating the vertical displacement of a particle chosen 
at the beginning and bottom of the spicule. We ignore the horizontal 
movement for this calculation. The figure shows that the particle 
follows the trajectory of the spicule quite well. The differences are 
caused by not considering horizontal displacement, approximations of 
the initial point of the spicule, and the numerical error of the 
displacement. 
For illustration purposes we also draw a trace of 
a signal propagating from the root of the spicule with 
the sound speed (blue). 

Figures~\ref{fig:traycl}-\ref{fig:traycr} illustrate the physics of 
spicule-like jets driven by a collapsing granule, magnetic energy 
release in the photosphere, and magnetic energy release in the 
chromosphere. We show as a function of time and height a range of 
parameters that play an important role in the momentum and energy 
balance of the atmosphere. From top to bottom and left to right: 
relative pressure perturbation $P/\mathrm{<}P\mathrm{>}-1$, vertical pressure gradient 
perturbation $(\partial P/\partial z)/\mathrm{<}\partial P/\partial z\mathrm{>}-1$, 
gravity perturbation, {\it i.e.}, the buoyancy force perturbation 
$g\rho/\mathrm{<}\partial P/ \partial z\mathrm{>} -1$, vertical Lorentz force 
perturbation $F_{Lz}/\mathrm{<}\partial P/ \partial z\mathrm{>}-1)$, the time 
derivative of the gas pressure per unit of mass $(1/\rho)\partial 
P/\partial t$, Joule heating per unit of volume normalized with   
the square root of the gas pressure 
$\rho q_{joule}/\sqrt{P}$, heat advection term per unit of mass $ - 
{\mathbf u}\cdot T \nabla s$, non-adiabatic heating per unit of mass  
(which we extract from the time derivative of the gas pressure by 
subtracting the adiabatic advection contribution), and the adiabatic 
advection term per unit of mass $(1/\rho)c_s^2 \nabla \cdot (\rho 
{\mathbf u})$. The brackets $\mathrm{<} \cdot \cdot \cdot \mathrm{>}$ denote temporal 
averaging, $\Gamma_3-1=(\partial \ln T/\partial \ln \rho)_s=(\partial 
P/\partial e)_{\rho}$ and $\rho$, ${\mathbf u}$, $e$, $T$, $s$, $c_s$ and $g$ are the density, 
velocity field, specific internal energy, temperature, specific entropy, sound speed and  
the constant of gravity acceleration, respectively. 

The pressure gradient perturbation imposes an upwards force when it is 
white, whereas the gravity perturbation imparts a downward force when 
it is white in figures~\ref{fig:traycl}-\ref{fig:traycr}. In other 
words, when both the pressure gradient and the gravity perturbation 
are white, then these forces are counteracting one another. 

The middle and bottom-middle images and last column of the 
figures~\ref{fig:traycl}-\ref{fig:traycr} show various terms 
contributing to the energy conservation equation. We can write the 
time derivative of pressure as:

\begin{eqnarray}
\frac{\partial P}{\partial t} & = & - c_s^2\nabla \cdot(\rho {\mathbf u}) + (\Gamma_3 -1)\rho(q_{joule}+q_{rad}+q_{visc} +q_{\kappa} - {\mathbf u} \cdot T \nabla s)
\end{eqnarray}

\noindent where $q_{joule}$, $q_{rad}$, $q_{visc}$, $q_{\kappa}$ are 
the specific heating terms from Joule dissipation, radiation, 
viscosity and thermal conduction, respectively. The Joule heating is 
defined as $q_{joule}=J^2/\sigma$, where $\sigma$ is the electrical 
conductivity. The specific heating by conduction is defined in the 
code as $q_{\kappa}=-\nabla(\kappa_{||}\nabla_{||}T)$, where the 
conduction is along the field lines (see Paper I for details). $\kappa_{||}$
is the conduction coefficient parallel to the field lines. The first term on the right hand 
side is the adiabatic contribution from the mass flux divergence. The 
second term is the non-adiabatic contribution arising from an 
imbalance between heating/cooling in a volume and the advection of 
heat into or out of the volume. 

\subsubsection{Spicule dynamics}
\label{sec:prof}

All the spicules we study here are characterized by a parabolic 
evolution with time of the top of the spicule, {\it i.e.}, the transition 
region. This is illustrated in Fig.~\ref{fig:traytg}, which shows the 
logarithmic temperature evolution of the plasma. The vertical velocity 
profile in this figure shows that the plasma in the spicule first 
moves upwards (white), followed by downward motion (black). 
The shock that drives the plasma upward is seen to propagate upwards 
with supersonic speed. The vertical acceleration panel shows upwards 
acceleration just before and below the spicule (white). Once the shock 
reaches the transition region, the plasma in the spicule shows 
deceleration (black) until the end of the spicule's life. This 
scenario is very similar to that found in the 1D and 2D simulations of 
\citet{Heggland:2007jt} and 
\citet{Hansteen+DePontieu2006}, respectively. 

Let us now study the momentum balance in more detail. The buoyancy force, 
pressure gradient force, and Lorentz force are all normalized with the 
temporal average of the pressure gradient, so that the figures can be 
compared directly. At the beginning stages of the spicule's evolution, 
there is typically a gas overpressure (white color, panel {\it a)} of 
figures~\ref{fig:traycl}-\ref{fig:traycr}). Before the spicule reaches 
its maximum height, this changes and becomes an underpressure (black 
color, panel {\it a)} of 
figures~\ref{fig:traycl}-\ref{fig:traycr}). While this pressure pulse 
reaches the transition region, we find that it originates in the lower 
chromosphere or photosphere depending on the driver. We will discuss 
the differences between each driver later. The pressure perturbation 
leads to an enhanced pressure gradient. The pressure gradient that 
originates in the driving region increases which leads to greater upflow 
velocities in the spicule. The gravity force increases (white), {\it i.e.}, 
dense plasma is lifted upwards, but it is not until the perturbation 
reaches the transition region that the overdense plasma is sufficient to 
compensate for the higher pressure gradient. The overdensity remains 
almost until the end of the spicule's evolution. The vertical Lorentz 
force is negligible in the evolution of the spicule compared to the 
pressure gradient and buoyancy force. 

Just at the top of the spicule, in the boundary between the 
transition region and the body of the spicule, the temporal variation 
of the pressure (panel {\it e)} of Figs.~\ref{fig:traycl}-\ref{fig:traycr})  
is negative (black). However, in the lower 
parts of the spicule, the variation of the pressure in time is 
positive (white). This increase in pressure follows the spicule trajectory 
when the spicule is increasing in length. Once the spicule reaches its 
maximum length, the variation of pressure in time is negative (black) 
and it follows the spicule trajectory back to lower heights. Joule 
heating (panel {\it f)} of Figs.~\ref{fig:traycl}-\ref{fig:traycr}) is not 
important inside the spicule, though it appears that in some 
cases Joule heating could play a role above the spicule. The 
advection term (panel {\it g)} of Figs.~\ref{fig:traycl}-\ref{fig:traycr}) is 
cooling (black) before, but heating after the spicule 
reaches its greatest length. Before the spicule reaches its maximum 
extent, cooling (black) by non-adiabatic advection 
(panel {\it h)} of Figs.~\ref{fig:traycl}-\ref{fig:traycr})  is strongest in the 
transition region and in a region extending 500~km downwards from the 
transition region. After reaching maximum extent, heating by 
non-adiabatic advection occurs in the same region. The advection term  
(panel {\it i)} of Figs.~\ref{fig:traycl}-\ref{fig:traycr}) 
is more important than the non-adiabatic 
contribution in the body of the 
spicule. The non-adiabatic term has the opposite sign of the adiabatic 
term. In other words, a positive contribution from the non-adiabatic 
term can balance the work done by adiabatic expansion ({\it i.e.}, a 
negative adiabatic contribution) and/or increase the pressure. 

\subsubsection{The dynamics of the driving mechanism}

\begin{figure}
	\includegraphics[width=17cm]{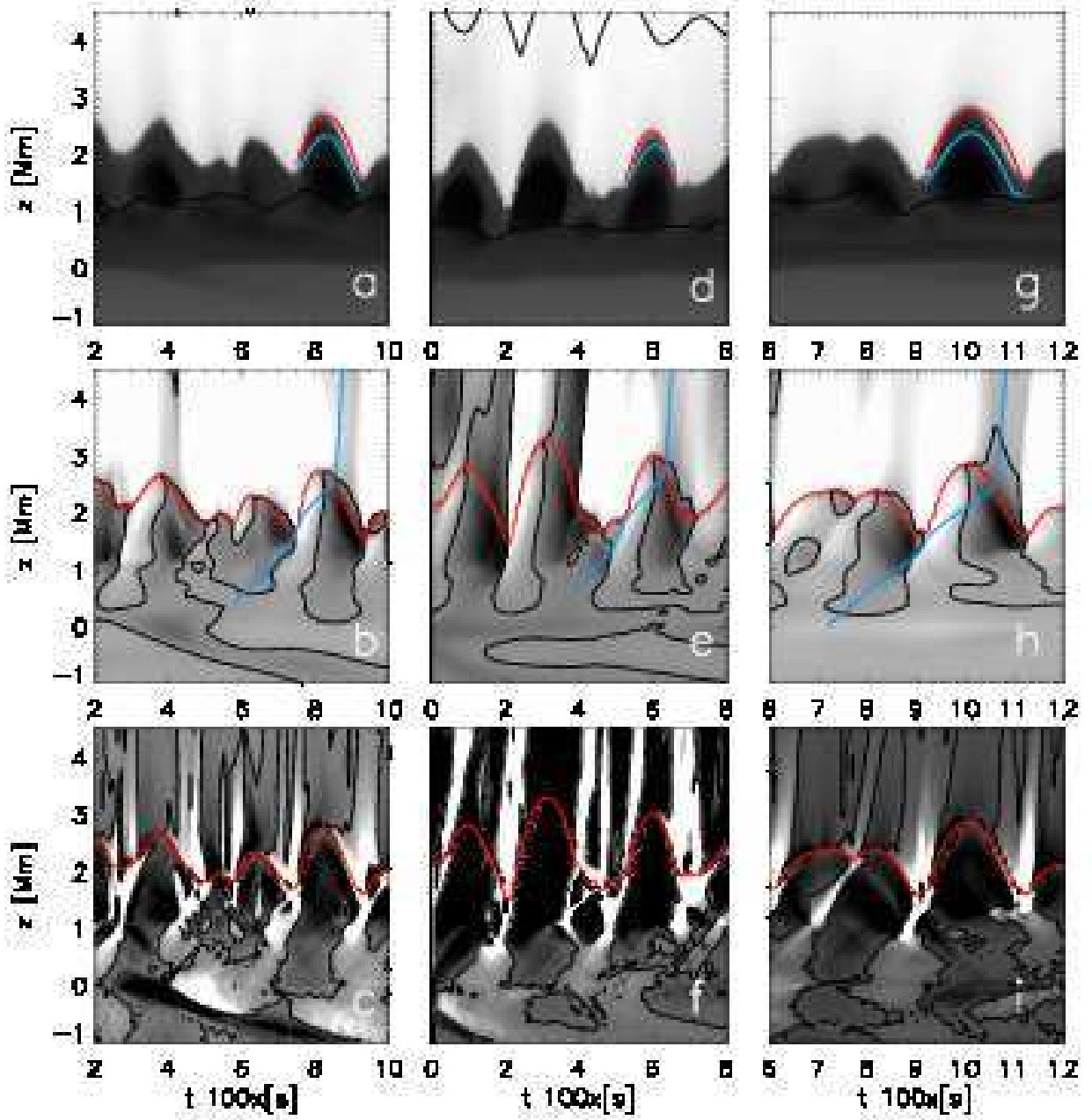} 
        \caption{\label{fig:traytg} Logarithmic temperature, vertical 
          velocity and vertical acceleration as a function time and of 
          height is shown from top to bottom row respectively. The 
          black contour in the first row is $\beta=1$, in the other 
          images it shows where the plotted parameter is zero. In the 
          first row, the parabolic fit is plotted with a red line, and 
          the particle trajectory with a blue line. The second and 
          third row show the transition region ($T=10^5$~K) as a red 
          line.  In the middle row, the blue line shows the path a 
          perturbation propagating at the speed of sound would 
          take. 
          The first column shows a spicule 
          that starts at $750$~s and that is driven by a collapsing 
          granule at time $400$~s.  The second column shows a spicule 
          that starts at $550$~s and that is produced by a 
          photospheric magnetic energy release event at time 
          $400$~s. The third column shows the evolution of a spicule 
          that starts at $950$~s and that is driven by a chromospheric 
          magnetic energy release at time 850~s.} 
\end{figure}

\begin{figure}
	\includegraphics[width=15.5cm]{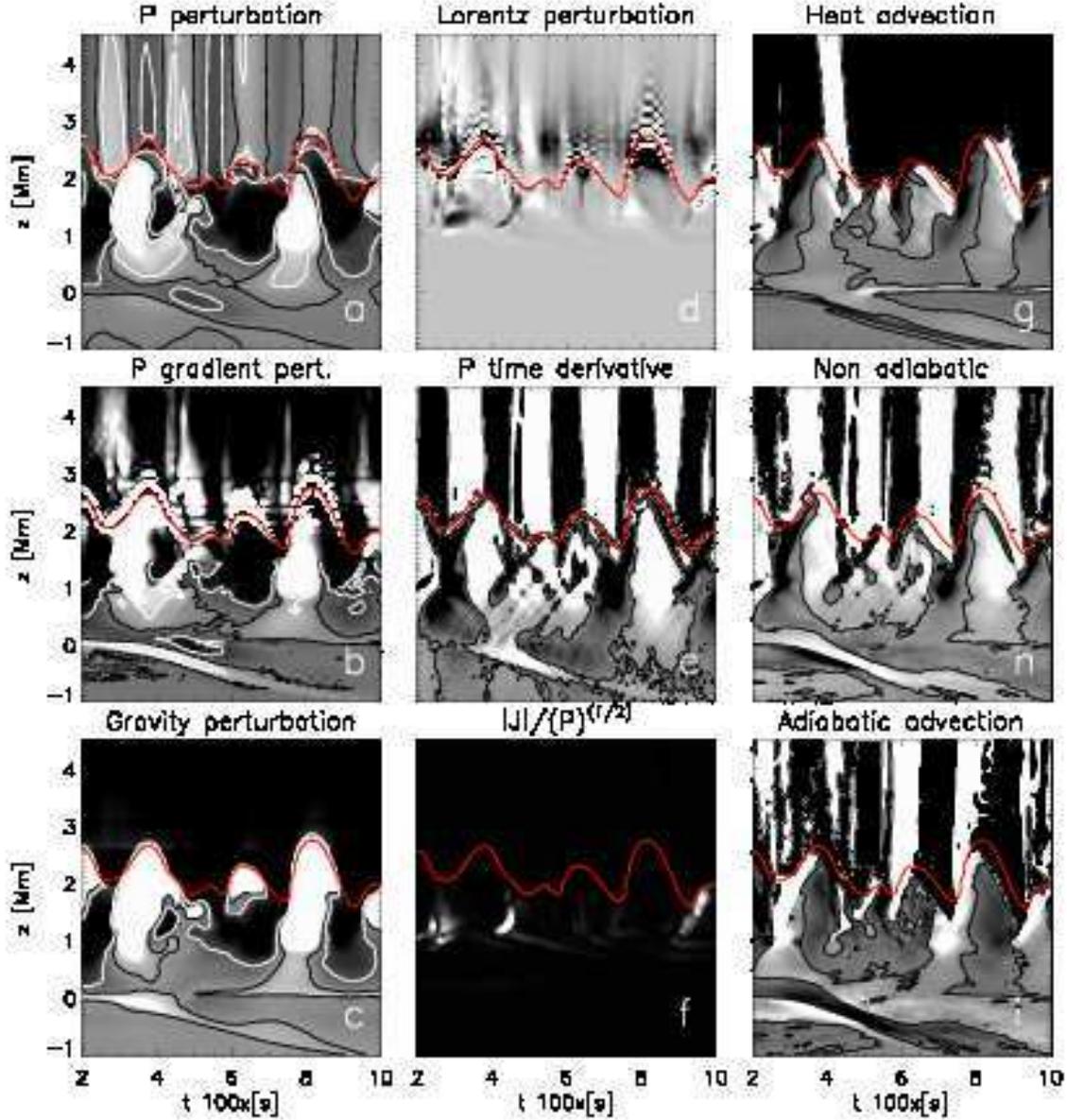}
        \caption{\label{fig:traycl} The following parameters are shown  
          from top to bottom and from left to right as  
          functions of height and time: 
          Pressure perturbation (white is 
          overpressure), pressure gradient perturbation (white is 
          upwards force), gravity perturbation (white is downwards 
          force), vertical Lorentz force perturbation (white is 
          upwards force), time derivative of pressure (white is 
          increase in pressure), $R_d$ term, see Eq.~\ref{eq:rd}  (white is large 
          magnetic discontinuity), heat advection term (white is large 
          contribution), non-adiabatic contribution (white is heating) 
          and adiabatic contribution (white is heating). The black 
          contour shows when the parameter is zero. The white contours 
          show perturbation of $\pm 10$\%. The red line is the 
          transition region ($T=10^5$~K). The example shown here is 
          the same spicule that is shown in the first column of 
          Fig.~\ref{fig:traytg}, which is driven by a collapsing 
          granule. The checkerboard pattern observed mostly in the 
          transition region in some panels (e.g., the Lorentz perturbation), 
          is caused by the numerical noise and spatial and temporal resolution in 
          the transition region and corona.}
\end{figure}

Spicule dynamics are the same irrespective of the driving 
mechanism, as can be seen in the different columns of 
Fig.~\ref{fig:traytg}.  However, in the convection zone, photosphere, 
and lower chromosphere the dynamics are different for the driving 
mechanisms we consider here, {\it i.e.}, collapsing granules, magnetic 
energy release in the photosphere, and magnetic energy release in the 
chromosphere. 

\citet{Skartlien:2000lr} have already described in detail the dynamics of 
collapsing granules and how these events can perturb the 
chromosphere. The results we find here are similar to those of 
\citet{Skartlien:2000lr}: We find that the photospheric perturbation 
associated with collapsing granules can 
perturb the chromosphere and produce a spicule.  
In short, in the convection zone the vertical mass 
flux divergence is compensated by horizontal mass flux convergence of 
fluid from neighboring granules. The collapsing granule undergoes a 
strong downward acceleration (see the dark feature which starts at 
400~s and $z\sim 0$~Mm panel {\it c)} of figure~\ref{fig:traytg}). In the upper 
photosphere ($z\approx 200$~km), the fluid is initially (300~s) accelerated 
downwards, but at $t\approx 400$~s upwards, which produces a 
rarefaction followed by a compression and then decaying oscillatory 
wake. This wake becomes a shock, around $t\approx 750$~s, in the 
transition region ($z\approx1800$~km) with velocities of the order of 
30~km~s$^{-1}$. 

When the granule collapses ($\sim 400$~s, $z\sim 0$~Mm), the adiabatic pressure 
(panel {\it i)} of Fig.~\ref{fig:traycl}) 
changes from a negative (black) to a positive (white) contribution, and the time 
derivative of the pressure (panel {\it e)} of Fig.~\ref{fig:traycl}) 
becomes positive (white). This perturbation 
propagates to higher and lower layers but it is largely compensated by the 
non-adiabatic contribution (panel {\it h)} of Fig.~\ref{fig:traycl}). Advective heating  
(panel {\it g)} of Fig.~\ref{fig:traycl}) dominates the non-adiabatic processes  
in a region from the lower chromosphere up to 
near the transition region ($z\approx [1-2]$~Mm). 

\begin{figure}
	\includegraphics[width=17cm]{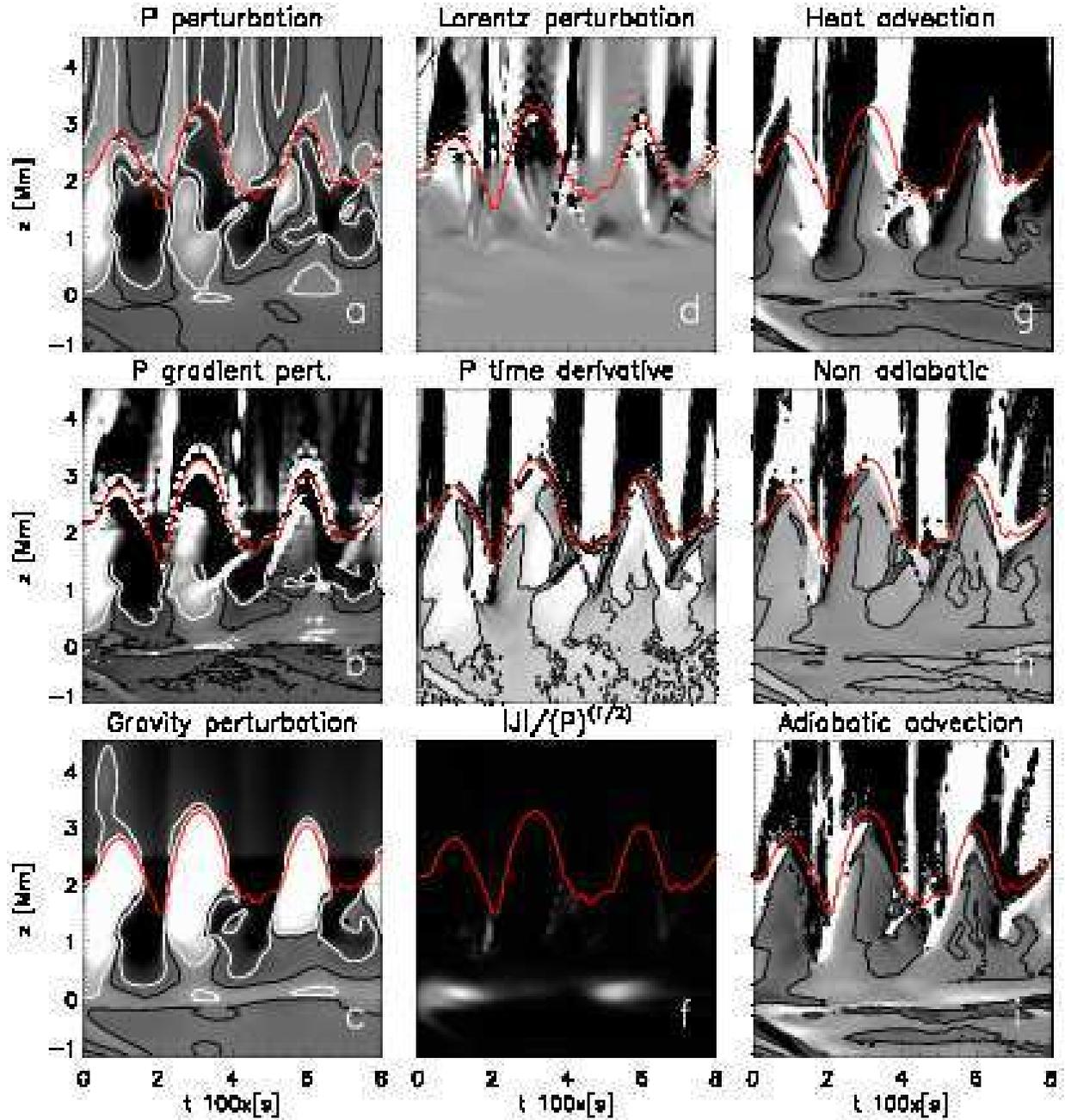}
        \caption{\label{fig:traycp} 
	Same terms, parameters and color scheme as described in 
	figure~\ref{fig:traycl} in the same order. 
	The example shown here is 
          the same spicule that is shown in the second column of 
          figure~\ref{fig:traytg}, which is driven by photospheric 
          magnetic energy release.} 
\end{figure}

\begin{figure}
	\includegraphics[width=17cm]{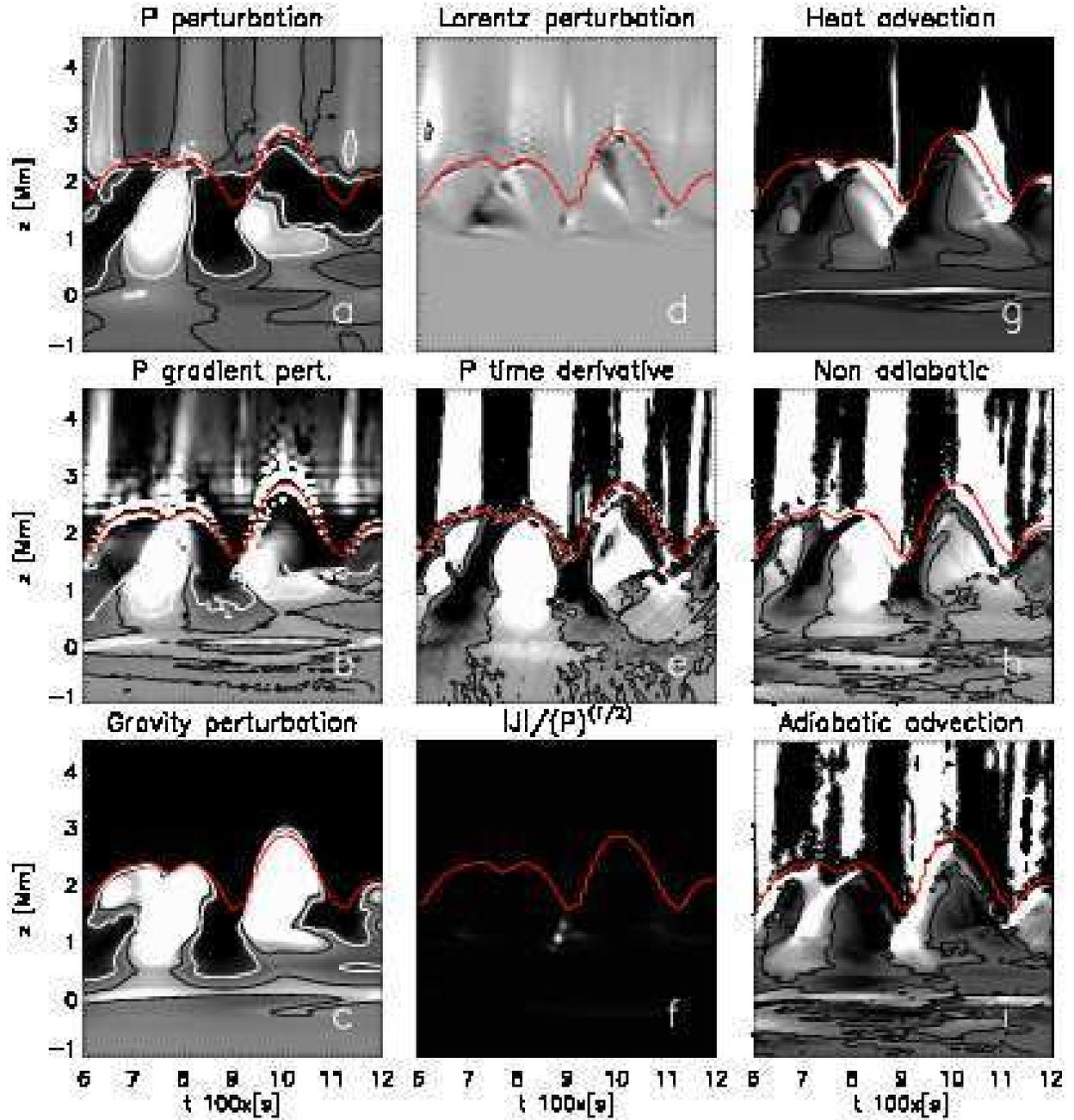}
        \caption{\label{fig:traycr} 
	Same terms, parameters and color scheme as described in  
	figure~\ref{fig:traycl} in the same order.  
	That correspond a spicule driven by a 
          chromospheric magnetic energy release shown in the third 
          column of figure~\ref{fig:traytg}.} 
\end{figure}

Figure~\ref{fig:traycp} and the middle column of figure~\ref{fig:traytg} show an example 
of a spicule driven by photospheric magnetic energy release which can be seen 
in the panel {\it f)} of Fig.~\ref{fig:traycp} during a period of time from   
50 to 200~s. The energy release is slightly extended in height and time.  
When a spicule is driven by photospheric magnetic energy release, an  
upwards acceleration in the upper photosphere is produced (shown in 
white in the panel {\it f)} of Fig.~\ref{fig:traytg} around 
$t\approx 350$~s and $z\sim 0$~Mm). As a result of the upward acceleration, the 
dynamics above the photosphere are similar to those we find for 
collapsing granules. However, in the photosphere and below the 
location of the driver, the dynamics are different. There is a small 
downward acceleration for a short time period (dark color around 
$z\approx[(-500)-0]$~km at time $t\approx[320-400]$~s at  
panel {\it f)} of Fig.~\ref{fig:traytg}). An overpressure appears when the 
wave propagates into the chromosphere (white color in the 
panel {\it a)} of Fig.~\ref{fig:traycp} at roughly $z\approx1000$~km 
and $t\approx 550$~s), as well as an excess in pressure gradient 
(white color of the panel {\it b)} of Fig.~\ref{fig:traycp} at a 
similar time and height as the overpressure).  We find that an excess 
in density appears over a larger region in time and height, which is 
opposite to the change in pressure gradient and a bit higher in amplitude (but a 
similar order of magnitude) when the wave gets closer to the 
chromosphere. In addition to this, we note that there is no adiabatic 
(panel {\it i)} of Fig.~\ref{fig:traycp}) nor 
non-adiabatic (panel {\it h)} of Fig.~\ref{fig:traycp}) 
contribution in the convection zone related with this 
type of driver: the convection does not 
seem to notice the magnetic energy release event. 

Let us now focus on chromospheric magnetic energy release (see  
the example shown in the last column of figure~\ref{fig:traytg} and 
figure~\ref{fig:traycr}). In that example the magnetic energy release is 
located in a narrow range in time and space,  at a height of 1~Mm 
above the photosphere and at $t\approx900$~s as shown in the 
panel {\it f))} of figure~\ref{fig:traycr}. We 
again find that above the magnetic energy release site the dynamics 
are similar to what occurs for the other drivers. However, below the magnetic energy 
release (lower chromosphere and photosphere) the dynamics are 
different. The magnetic energy release produces a downward 
acceleration (in darkish color starting at $t\approx900$~s and at 
$z\approx 1$~Mm panel {\it i)} of Fig.~\ref{fig:traytg}) which 
propagates towards the photosphere. The rarefaction and subsequent 
compression are nearly in phase below the magnetic energy release 
height (see panels {\it g)} and {\it h)} of 
Fig.~\ref{fig:traycr} at times $t>900$~s).  An excess in pressure 
gradient (panel {\it b)} of Fig.~\ref{fig:traycr}), balanced by an excess in density  
(panel {\it c)} of Fig.~\ref{fig:traycr}), appears when and where the 
magnetic energy release takes place (around $900~s$ and at $z\approx 
1$~Mm). These increases in density, pressure, and pressure gradient move 
with the spicule until it reaches its maximum extent, but also remain for a few 
minutes at the height ($z\approx1$~Mm) where the magnetic energy 
release takes place. However, these increases do not extend down to below the 
photosphere.

For both types of spicule driven by magnetic energy release, the adiabatic 
pressure increases just above the location of the driver and the time 
derivative of pressure becomes positive and propagates to higher and 
lower layers ($z\approx0$~km and $t\approx320$~s for the photospheric 
magnetic energy release shown in Fig.~\ref{fig:traycp} and 
$z\approx1$~Mm and $t\approx900$~s for the chromospheric magnetic 
energy release shown in Fig.~\ref{fig:traycr}). This is counteracted 
by the non-adiabatic contribution. However, this changes rapidly in 
higher layers, close to the transition region, where we find a growing 
adiabatic contribution. Below roughly $z\approx1.5$~Mm the  
adiabatic contribution is compensated by non-adiabatic contribution.  
At greater heights the advective heating dominates the 
non-adiabatic contribution, {\it i.e.} in 
the region confined between the lower chromosphere to near the 
transition region (see panel {\it g)} of 
Fig.~\ref{fig:traycp}~and~\ref{fig:traycr}). 

\section{Discussion and conclusions}
\label{sec:conclusions}

Our 3D simulations naturally produce structures that resemble observed type~{\sc i} 
spicules. These models represent a significant improvement over the 1D and 2D 
simulations that have been used to study these kinds of spicules in the 
recent past \citep{Heggland:2007jt,Hansteen+DePontieu2006}, since they 
are the most realistic 3D radiative MHD simulations to date. Similar to the 2D simulations, the current
simulations include non-grey and non-LTE radiative transfer with scattering, thermal 
conduction along the field lines, and a convectively unstable convection zone. The current simulations
also include a full 3D approach, a corona that is maintained self-consistently with 
temperatures greater than $5\,10^5$~K, and flux emergence. 

In total we found 168 spicules of which 150 follow a clear parabolic 
trajectory in which the transition region is first pushed upwards by 
denser chromospheric plasma, and then retreats 
back during the second half of the spicule's life. In analyzing these 
spicules we follow the approach taken by 
\citet{Hansteen+DePontieu2006} and \citet{De-Pontieu:2007cr}: we 
determine the deceleration, maximum 
velocity, duration and maximum height from parabolic fits. We find that the ranges of 
values for deceleration and maximum velocity are similar to those 
found for spicular jets in previous simulations, and for spicules, 
mottles and fibrils found in observations. 

The ranges of values for duration and maximum height are smaller 
than those found in observations and previous simulations. All the 
jets we find here have durations of order 2-3 minutes, whereas 
observations show a wider variety of durations, from 2 to 8 minutes. In 
addition, we find rather short jets in the 3D simulations, with lengths 
typically less than 2000 km. It is likely that a combination of 
limitations of the numerical simulations is the cause for these 
discrepancies. For example, the duration and shorter lengths of the 
simulated jets are actually quite similar to those found in 
observations of dynamic fibrils that occur in a dense plage region 
where the magnetic field is more vertical. It is thus 
likely that the limited range of magnetic field configurations in our 
simulations as compared to the variety of magnetic environments that 
the real Sun presents plays a role in this discrepancy. 

Additionally, our selection criterion imposes a minimum height of 2000 
km on the spicules. This provides a bias towards vertical features, 
since it potentially removes candidate spicules that are heavily 
inclined and that form along the magnetic canopy. These inclined 
spicules may not cause much of a transition region height excursion, 
but may well be an important fraction of what we observe on the disk 
on the Sun as indicated by the heavily inclined dynamic fibrils that 
dominate the less dense plage region discussed by 
\citep{De-Pontieu:2007cr}. These inclined spicules are also seen to form 
regularly in previous 2D simulations \citep{Hansteen+DePontieu2006}. 
Another bias that leads to vertically oriented spicules in this paper is that the  
field lines that extend into the corona, {\it i.e.} the field lines on which  
spicules occur, are essentially vertical in the magnetic field topology 
modeled in both models. Field lines that  
close lower in the chromosphere may hinder the formation of  
spicules through the processes of wave mode conversion and reflection as 
described by {\it e.g.} \citet{Rosenthal:2002mw} and \citet{bogdan2003}. 

One of the mechanisms that is thought to drive a significant fraction 
of spicule-like jets in the atmosphere \citep{De-Pontieu:2004hq}, 
{\it i.e.}, leakage of p-mode oscillations, likely play a significant role  
in the group ``other mechanisms'', which constitutes 26\% of 
our spicules in both simulations. For half of these spicules, we find a  
correspondence between the timing of p-modes oscillations in the 
photosphere and the initiation of the 
chromospheric waves. The p-modes in the simulations 
are stochastically excited by the convective motions but the lower 
cavity boundary 
is given by the simulation box rather than by refraction in the 
deep convection zone. The resulting modes are therefore rather 
few but the total power is similar to the solar case. At $z=200$~km 
we have an rms velocity amplitude of 260~m~s$^{-1}$ in the oscillation  
component (horizontal phase speed larger than 6 km~s$^{-1}$) below 
a frequency of 4.5 mHz. This value is of the same order as the observed rms 
velocity amplitude of 
202~m~s$^{-1}$ in MDI high-resolution data \citep{Straus+etal1999} 
that is filtered in the same way  
(Straus, T.: 2009, personal communication). The smaller number of 
modes present in the simulations compared with the Sun may, however,  
lead to fewer occurrences of large amplitude, constructive interference 
events. In addition, propagation of p-mode oscillations into the  
chromosphere (and subsequent spicule formation) are  
facilitated by the presence of inclined magnetic field lines, which are rare in the 
magnetic field configurations in our models. These two factors may help explain the 
relatively low occurrence rate of p-mode driven spicules in the simulations we analyze here. 

Another factor that could reduce the strength of the upwardly propagating 
shocks in our simulations compared to the Sun is the treatment of hydrogen  
ionization. In our simulation it is implicitly assumed that hydrogen ionizes or 
recombines immediately as a result of changes in the temperature and density. 
\citet{Carlsson:2002wl,Leenaarts+etal2007} and \citet{Leenaarts:2007sf}  
have shown that this is not the  
case and that a proper treatment of hydrogen ionization can have a significant 
impact on shock characteristics in the upper chromosphere. This is because the energy 
going into ionization in the current simulations would be available for heating the  
plasma and thus strengthening the shocks. 

Finally, it is likely that the low 
numerical resolution in the chromosphere and transition region in both 
the horizontal and vertical direction play an important role in the 
discrepancy between jets in the simulations and observed jets. The resulting sizable 
numerical diffusion can significantly reduce the strengths of shocks, 
which can lead to jets of reduced amplitude (and thus reduced 
duration, since we impose a minimum height of 2000 km). We expect that 
increasing the number of grid points (and thus decreasing the 
viscosity and diffusion) of the computational domain can lead to significant 
differences in the properties of the spicules, especially with respect to 
the height and duration. 

Despite all of these differences, we find many similarities 
between observations and our simulations with respect to the 
correlations between various 
parameters. \citet{Hansteen+DePontieu2006} and \citet{De-Pontieu:2007cr} find a host of 
correlations between the deceleration, maximum velocity, length and 
duration of dynamic fibrils. The spicule-like features in our 
simulations reproduce these correlations remarkably well. This is 
because the underlying cause for these correlations is shock-wave 
physics, as discussed at length by \citet{Heggland:2007jt}. Our 
simulations here show in detail how the parabolic trajectory of the 
spicules is caused by upper chromospheric flows driven by a shock wave 
propagating upward after being generated below the spicule. Detailed 
analysis of the momentum and energy balance in our 3D simulations 
shows that the pressure gradient resulting from the shock wave passage 
counteracts the gravitational force. This naturally leads to a 
trajectory that is not a purely ballistic trajectory: the shock wave 
and its associated pressure gradient provide an ongoing force against 
gravity. As a result, we find decelerations that are substantially 
different from solar gravity (similar to what was found in previous 
simulations and the observations). The spicule trajectory is a natural 
result of the motion induced by the upflows and downflows in the wake 
of the shock wave that ultimately propagates into the corona. In 
addition, the non-adiabatic contribution to the energy balance 
counteracts the adiabatic contribution during the early evolution of the 
spicule. 

The type~{\sc i} spicules in our simulations (and previous 
simulations) are formed by upwardly propagating chromospheric waves 
that form into shocks. In our simulations, we manage to identify 
several different driving mechanisms: collapsing granules, magnetic 
energy release in the photosphere and magnetic energy release in the 
low chromosphere. We also identified other candidates, such as  
p-modes,  
convective buffeting by breaking granules, and magnetic field 
topology changes resulting from flux emergence. This list is not 
exhaustive, since our simulations clearly show that {\it any 
  mechanism} that produces a chromospheric wave or perturbation that 
develops into a shock and propagates along a flux concentration can 
drive type~{\sc i} spicules. It is thus likely that there are 
more/other physical processes that can produce the waves that drive 
type~{\sc i} spicules. Given the complexities of the 3D simulations, 
it is generally quite difficult to pinpoint the driving mechanism for 
the waves that drive each spicule. We find that as long as a 
perturbing event creates a perturbation close to a flux concentration 
that reaches into the chromosphere and corona, a spicule can 
develop. The properties of the spicules do not seem to change much for 
the various driving mechanisms that we identified, except perhaps when 
magnetic energy is released in the chromosphere. In that case, there 
seems to be a slightly different relation between deceleration and 
maximum velocity, which is concentrated in the upper part of the 
distribution of points (see Fig.~\ref{fig:dist}). 

The magnetic energy release events described here involve regions 
where the current is highly localized and Joule dissipation is locally 
significantly enhanced, with a subsequent perturbation propagating 
away from the dissipation site. Some of these events might be related 
to magnetic field line reconnection, although in most cases 
reconnection of field lines is difficult to identify. Our simulations 
suggest \citep[see also][]{Heggland:2009lq} that, in principle, 
sudden energy release ({\it e.g.}, from reconnection) in the low atmosphere 
can lead to both type~{\sc i} and type~{\sc ii} jets. Our simulations 
contain a few events that are more violent with shorter lifetimes and 
higher velocities that are similar to type~{\sc ii} 
spicules. These jets will be described in future work. It is clear 
however that the underlying formation mechanism and/or magnetic 
configuration is different between these two types of events. 

The importance of the magnetic field configuration in producing  
type~{\sc i} spicules is significant. While our simulations do not cover a 
wide range of field configurations (unipolar plage, mixed polarity 
quiet Sun, open field in coronal holes), we already notice from our two 
different simulations that the number of spicules increases with the strength of the ambient 
field and that they become shorter, which is similar to what is 
observed on the Sun. The most important impact of field configuration 
entry into the corona 
lies in the different driving mechanisms that are expected to dominate 
in different field topologies. For example, our simulations involve 
emergence of an intense magnetic flux tube. It is thus not surprising 
that about half of the events we find are driven by magnetic energy 
release, most likely related to the emergence of new field into a 
pre-existing ambient field. Such conditions are not typical for plage 
regions or most quiet Sun regions, so that we should be careful in 
extrapolating the importance of the different mechanisms to the real 
Sun. One of the two driving mechanisms that we expect to occur 
everywhere on the Sun, {\it i.e.}, convective buffeting of flux 
concentrations, is shown to play an important role in producing 
spicules in our simulations, and can be expected to play a similar 
role on the Sun.  

\section{Acknowledgments}

This research has been supported by a Marie Curie Early Stage Research 
Training Fellowship of the European Community's Sixth Framework 
Programme under contract number MEST-CT-2005-020395: The USO-SP 
International School for Solar Physics. Financial support by the 
European Commission through the SOLAIRE Network (MTRN-CT-2006-035484) 
is gratefully acknowledged. The 3D simulations have been 
run with the Njord and Stallo cluster from the Notur project and the 
Columbia cluster of NASA's High-End Computing Program. We thankfully 
acknowledge the computer and supercomputer resources by the Research 
Council of Norway through grant 170935/V30 and through grants of 
computing time from the Programme for Supercomputing. BDP was 
supported by NASA grants NNM07AA01C (HINODE), NNG06GG79G and 
NNX08AH45G. To analyze the data we have used IDL and Vapor 
(http://www.vapor.ucar.edu).

\begin{deluxetable}{ccccccccc}
\tablecaption{\label{tab:runs} Summary of the parameters which describe the two simulations. From left to right, the assigned name, the twist of the tube, the magnetic field strength of the tube, the size of the computational box, the duration of the simulations, the radius of the tube and the ambient field measured at the photosphere.}
\tablehead{
\colhead{Name} & \colhead{Twist $\lambda$}  & \colhead{B$_0$ [G]} & \colhead{Size [$Mm^3$]}  & \colhead{Time [s]} & \colhead{Radius [Mm]} & \colhead{Ambient field [G]}}
\startdata
A2 & 0.6 & 4484 & $16\times 8\times 16$ &  5700 &  0.5 & 16 \\ \\
B1 & 0 & 1121 & $16\times 8\times 16$ &  4400 &  1.5 & 160 \\ \\
\enddata
\end{deluxetable}

\end{document}